\documentclass[aps,pra,preprint,groupedaddress,showpacs,showkeys,preprintnumbers]{revtex4-1}
\RequirePackage{times}
\RequirePackage{mathptm}
\usepackage{xcolor}
\usepackage{graphicx}
\usepackage{amssymb}
\usepackage[latin1]{inputenc}
\usepackage{array}
\usepackage{longtable}
\usepackage{calc}
\usepackage{multirow}
\usepackage{hhline}
\usepackage{ifthen}
\usepackage{color}

\begin{document}

\title{How Random Is Quantum Randomness? \\ An Experimental Approach}

\author{Cristian S. Calude}
\email{cristian@cs.auckland.ac.nz}
\homepage[]{http://www.cs.auckland.ac.nz/~cristian}

\author{Michael J. Dinneen}
\email{mjd@cs.auckland.ac.nz}
\homepage[]{http://www.cs.auckland.ac.nz/~mjd}

\affiliation{Department of Computer Science, University of Auckland,
Private Bag 92019, Auckland, New Zealand}

\author{Monica Dumitrescu}
\affiliation{Faculty of Mathematics and Computer Science, University of Bucharest,
Str. Academiei 14, 010014 Bucharest, Romania}
\email{mdumi@fmi.unibuc.ro}
\homepage[]{http://fmi.unibuc.ro/ro/dumitrescu_monica}

\author{Karl Svozil}
\affiliation{Institute for Theoretical Physics, University of Technology Vienna,
Wiedner Hauptstrasse 8-10/136, 1040 Vienna,  Austria}
\email{svozil@tuwien.ac.at}
\homepage[]{http://tph.tuwien.ac.at/~svozil}

\date{\today}

\begin{abstract}
Our aim is to experimentally study the possibility of distinguishing between quantum sources of randomness---recently proved to be theoretically incomputable---and  some well-known computable sources of  pseudo-randomness. Incomputability is a necessary, but not sufficient ``symptom'' of ``true randomness.'' We base our experimental approach on algorithmic information theory which provides characterizations of algorithmic random sequences in terms of the degrees of incompressibility of their finite prefixes. Algorithmic random sequences are incomputable, but the converse implication is false. We have performed tests of randomness on pseudo-random strings (finite sequences) of length $2^{32}$ generated with software (Mathematica, Maple), which are cyclic (so, strongly computable), the bits of $\pi$, which is computable, but not cyclic, and strings produced by quantum measurements  (with the commercial device Quantis and  by the Vienna IQOQI group). Our empirical tests indicate quantitative differences, some statistically significant,  between computable and incomputable sources of ``randomness.''
\end{abstract}

\pacs{03.67.Lx, 05.40.-a, 03.65.Ta, 03.67.Ac, 03.65.Aa}
\keywords{quantum randomness, quantum indeterminism, random processes, quantum algorithms}
\preprint{CDMTCS preprint nr. 372/2009}
\maketitle

\section{Introduction}
\thispagestyle{empty}

From the 16th century onwards, following Galilei, Kepler, Leibniz, Newton and others,
the rise of determinism culminated around the time of the French and American Revolutions
with Laplace's research on the stability of the solar system without divine intervention~\cite{frank}.
In the late 19th century, first indications of potential limits to
the pure deterministic research program emerged, in particular with Poincar{\'{e}}'s contribution~\cite{poincare14,Diacu96} to the
solution of the three-~\cite{Sundman12} and general $n$-body problem~\cite{Wang91,Wang01,Diacu96},
which is often considered as a precursor of chaos theory~\cite{eckmann1,Diacu96-ce}.

Soon, and despite the reluctance and opposition of many of its creators,
most notably Planck~\cite{born-55}, Einstein~\footnote{
Recall Einstein's {\it dictum} in a letter to Born, dated December~12th, 1926~\cite[p.~113]{born-69},
``In any case I am convinced that he [[the Old One]] does not throw dice.''
(In German: ``Jedenfalls bin ich {\"{u}}berzeugt, dass der [[Alte]] nicht w{\"{u}}rfelt.'')},
Schr\"odinger and De Brogli,
quantum mechanics began to be accepted as an irreducibly probabilistic theory,
postulating an indispensable ``objective'' (in distinction to ``epistemic;'' cf. below) random behavior of individual particles,
while their probabilities follow deterministic laws.
With the rise of quantum mechanics (and later on also chaos theory), the {\em principle of sufficient reason}
--- stating that every phenomenon has its explanation and cause ---
had to be partially abandoned.
Indeed,  indeterminism and randomness in quantum mechanics,
as postulated by Born, Heisenberg, Bohr and Pauli~\cite[p.~115]{pauli-probaphysics}
is commonly believed, accepted and canonized to the extent  that~\cite{zeil-05_nature_ofQuantum}
``the discovery that individual events are
irreducibly random is probably one of the
most significant findings of the twentieth
century. [[$\ldots$]]~for the individual event in quantum physics, not only do we not know the cause, there is no cause.''

However, insufficient causation needs not be perceived merely negatively as a lack of prediction or control.
Today it is widely acknowledged that certified randomness can be a valuable resource
(e.g., for testing primality~\cite{ch-schw-78,Granville-92}), and that under various circumstances
a lack of randomness may have negative consequences (e.g., erroneous numerical calculations~\cite{PhysRevE.69.055702}).
The pitfalls of software-generated pseudo-randomness~\cite{v-neumann-50}
are well-known~\cite{Marsaglia-68,DBLP:journals/ibmrd/Pickover91,Bowman1995315,PhysRevE.69.055702}.
In John von Neumann's words~\cite{von-neumann1}:
``Anyone who considers arithmetical methods of producing random digits is, of course, in a state of sin.''

Classical physical processes are subject to difficulties with ``subjective''  or ``epistemic'' randomness
(a criticism often attributed to Heisenberg~\cite{zeil-05_nature_ofQuantum})
---
people consider events to be random when they cannot detect
any regularities characterizing the structure of those events,
yet the events {\em could} still be causally described if they
would  know enough about the evolution of the system
---
or even bias; the typical example being coin tosses~\cite{diaconis:211}.
Several methods to generate random sequences from physical processes have been proposed~\cite{csw:prg},
among them the coding of electric pulses~\cite{0022-3735-3-8-303},
or semiconductor devices~\cite{Agnew-87,Aware,Araneus,comscire,LavaRnd,dynes:031109,Ma:05,stipcevic045104,Wayne-09}.
The first book~\cite{rand-55} containing a  million  of random digits  using a physical source of randomness
was published by The RAND Corporation in  1955~\footnote{According to The RAND Corporation's disclosure,
``The random digits in this book were produced by re-randomization of a basic table generated
by an electronic roulette wheel. Briefly, a random frequency pulse source,
providing on the average about 100,000 pulses per second, was gated about once
per second by a constant frequency pulse. Pulse standardization circuits passed
the pulses through a 5-place binary counter. In principle the machine was a 32-place
roulette wheel which made, on the average, about 3000 revolutions per trial and produced
one number per second. A binary-to-decimal converter was used which converted 20 of the 32 numbers
(the other twelve were discarded) and retained only the final digit of two-digit numbers;
this final digit was fed into an IBM punch to produce finally a punched card table of random digits.''
}.

Currently there are two main sources capable of generating very fast large amounts of  ``random''
bits: software-generated randomness
(pseudo-randomness) and quantum randomness. Quantum randomness has been used as an ``objective'' resource of randomness
through various processes, in particular the decay of meta-stable states~\cite{PhysRevLett.54.1023,er-put:85,erber-95}
(for a criticism, see~\cite{knight-86}) or radioactive decays~\cite{schmidt:462,walker-hotbits},
arrival times~\cite{stipcevic4442,stipcevic045104,dynes:031109,Ma:05,Wayne-09},
or the passage through some beam splitter~\cite{svozil-qct,rarity-94,zeilinger:qct,stefanov-2000,0256-307X-21-10-027,wang:056107,fiorentino:032334,svozil-2009-howto,Kwon:09}.

How different are these sources? Recently it has been proved that quantum randomness is incomputable (see more details in  Section~\ref{incomput}). Incomputability is a necessary, but not sufficient ``symptom''
of ``true randomness.''  Can we
 experimentally  distinguish between
quantum  and   computable
sources of  ``randomness?''
In what follows, we   answer this question in the affirmative using an experimental approach based on algorithmic information theory
which provides characterizations of algorithmic random sequences in terms of the degrees of incompressibility of their finite prefixes. Algorithmic random sequences are incomputable, but the converse implication is false.

We have performed tests of randomness on pseudo-random strings (finite sequences) of length $2^{32}$ generated with software
(Mathematica, Maple), which are  cyclic (so, strongly computable), the bits of $\pi$, which is computable, but not cyclic, and strings produced by quantum measurements  (with the commercial device Quantis and  by the Vienna IQOQI group).

The paper is organized as follows. In the following section we present quantum randomness;
in Section~\ref{tests} we present the main tests and results; Section~\ref{conclusion} includes our conclusions.

\section{Quantum randomness}

In three distinct but intricately interlinked ways, the evolution of quantum mechanics ordained the abandonment
of absolute determinism, and has established a clearly defined mixture of determinism and indeterminism,
at least in the mainstream perception of the formalism~\cite{jammer:89,jammer1,feynman-law,fuchs-peres,clauser-talkvie}:
\renewcommand{\labelenumi}{(\roman{enumi})}
\begin{enumerate}
\item
random occurrence of individial events~\cite{born-26-1,born-26-2} or outcomes for quantized systems
which are in a superposition of eigenstates of the hermitean operator
corresponding to the observable;
i.e., randomness from projection measurements on superposition states;
\item
complementarity, as proposed by Pauli~\cite{pauli:58}, Heisenberg, Dirac and Bohr;
\item
value indefiniteness~\cite{peres222} as implied by the theorems of Bell, Kochen \& Specker and Greenberger, Horne \& Zeilinger~\cite{mermin-93}.
\end{enumerate}

\subsection{Random individual measurement outcomes}

With respect to the perception of certain individual outcomes of measurements,
the year 1926 marked the emergence of Born's acausal, indeterministic and
probabilistic interpretation of
Schr\"odinger's wave function as a complete and maximal description of a quantum mechanical state.
Born states that (cf.~\cite[p.~866]{born-26-1}, English translation in Ref.~\cite[p.~54]{wheeler-Zurek:83})~\footnote{
{ ``Vom Standpunkt unserer Quantenmechanik gibt es keine Gr\"o\ss e, die im {\em Einzelfalle} den Effekts eines Sto\ss es
kausal festlegt; aber auch in der Erfahrung haben wir keinen Anhaltspunkt daf\"ur, da\ss~ es innere Eigenschaften
der Atome gibt, die einen bestimmten Sto\ss erfolg bedingen.
Sollen wir hoffen, sp\"ater solche Eigenschaften
[[$\ldots$]] zu entdecken und im Einzelfalle zu bestimmen?
Oder sollen wir glauben, dass die \"Ubereinstimmung von Theorie und Erfahrung
in der Unf\"ahigkeit, Bedingungen f\"ur den kausalen Ablauf anzugeben, eine pr\"astabilisierte Harmonie ist,
die auf der Nichtexistenz solcher Bedingungen beruht?
Ich selber neige dazu,die Determiniertheit in der atomaren Welt aufzugeben.''
}
},
\begin{quote}
{  ``From the standpoint of our quantum mechanics, there is no quantity
which in any individual case causally fixes the consequence of the collision;
but also experimentally we have so far no reason to believe that there are some inner properties of the atom
which condition a definite outcome for the collision.
Ought we to hope later to discover such properties [[$\ldots$]]  and determine them in individual cases?
Or ought we to believe that the agreement of theory and experiment --- as to the impossibility
of prescribing conditions? I myself am inclined to give up determinism in the world of atoms.''
}
\end{quote}
While postulating a probabilistic behavior of individual particles,
Born offers a deterministic evolution of the wave function
(cf.~\cite[p.~804]{born-26-2}, English translation in Ref.~\cite[p.~302]{jammer:89})~\footnote{
{  ``Die Bewegung der Partikel folgt Wahrscheinlichkeitsgesetzen,
die Wahrscheinlichkeit selbst aber breitet sich im Einklang mit dem Kausalgesetz  aus.
[Das hei\ss t, da\ss~ die Kenntnis des Zustandes in allen Punkten in einem Augenblick
die Verteilung des Zustandes zu allen sp{\"a}teren Zeiten festlegt.]''
}
},
\begin{quote}
{  ``The motion of particles conforms to the laws of probability, but the probability itself
is propagated in accordance with the law of causality.
[This means that knowledge of a state in all points in a given time determines the distribution of
the state at all later times.]''
}
\end{quote}

At the time of writing this statement Born did not specify the formal notion of ``indeterminism'' he was relating to.
So far, no mathematical characterization of quantum randomness has been proven.
In the absence of any indication to the contrary, it is mostly implicitly assumed
that quantum randomness is of the strongest possible type;
which amounts to postulating that the associated sequences are algorithmically incompressible.
This does not exclude the possibility of weaker forms of randomness being generated by quantum measurements.

Random individual outcomes may occur at least in two different ways:
(i) either due to a context mismatch between preparation and measurement,
(ii) or
due to an ignorance of the state preparation resulting in a mixed state.
In what follows, we shall discuss these issues in some detail.

We shall consider normalized states.
The superscript ``$T$'' indicates transposition.
If not stated otherwise, we shall adopt the notation of Mermin's book on {\em Quantum Computer Science}
\cite{mermin-07}.
A quantum mechanical context~\cite{svozil-2008-ql}
is a ``maximal collection of co-measurable observables'' constituting
a ``classical mini-universe'' within the nondistributive structure of quantum propositions.
It can be formalized by a single  ``maximal'' self-adjoint operator.
Every collection of mutually compatible co-measurable operators (such as projections corresponding to yes--no propositions)
are functions of such a maximal operator
(e.g., Ref.~\cite[Sec.~II.10, p. 90, English translation p.~173]{v-neumann-49},
Ref.~\cite[\S~2]{kochen1}, Ref.~\cite[pp.~227,228]{neumark-54}, and Ref.~\cite[\S~84]{halmos-vs}).

\subsubsection{Mismatch between state preparation and measurement}
\label{2009-QvPR-s-mismatch}

There might be a {\em context} mismatch between state preparation and measurement;
i.e.,  the system has been prepared in a pure state
corresponding to a certain context (maximal observable),
and is measured in another,  complementary (see below) context
(maximal observable).
In such a case, the state of the system --- in terms of the spectral decomposition
of the measurement context --- is in a {\em coherent} superposition of at least some eigenstates of the preparation context.
An ``irreversible'' measurement~\cite{hkwz,greenberger2} ``reduces'' the state to one of the eigenstates
of the measurement context.
According to the Born rule (e.g.,~\cite[Chapter~1]{mermin-07}),
the probability of the occurrence of any such measurement outcome labelled by $i$
is given by the absolute square of the scalar products
$\vert \langle \psi_i \vert \varphi \rangle \vert^2$
between the state $\vert \varphi \rangle$  in which the system has been prepared
and the corresponding eigenstate $\vert \psi_i \rangle$
of the context.
Other than this probabilistic law, quantum mechanics renders no further prediction for the occurrence of single measurement
outcomes.
Note that the amount of indeterminacy (as measured by the lack of bias of measurement outcomes
formalizable in terms of average algorithmic information increase per outcome)
increases with the ``apartness'' of the preparation and measurement properties;
i.e., with the magnitude of the context mismatch.
On the average, conjugate bases~\cite[p.~86]{wiesner} assure the greatest context mismatch,
and hence the greatest degree of randomness gain per experiment.

Quantum realizations of the method have been proposed~\cite{svozil-qct,rarity-94}, patented~\footnote{
See also the later patents at Refs.~\cite{dultz-98,dultz-99},
as well as at Refs.~\cite{Ribordy-04,Ribordy-06}.}
and realized~\cite[Fig.~1(b)]{zeilinger:qct} (see also~\cite{stefanov-2000})
for a delayed choice Bell-type experiment~\cite{zeilinger-epr-98}.
Note that in the latter experimental realization, in the second {\em modus operandi} of~\cite{zeilinger-epr-98},
light of very low intensity --- the photon production rate should be much smaller than the corresponding coherence time ---
is prepared by sending it through a linear polarizer, e.g., in the vertical direction $\updownarrow$,
which guarantees that (ideally) only photons in a definite,
pure state corresponding to the polarization direction $\updownarrow$ leave the polarizer.
The photons impinge on a beam-splitting polarizer,
which should (ideally) be maximally (anti)aligned at exactly $45^\circ$ ($\pi/4$~radians) in order to yield a 50:50 ratio of photons
polarized in either one of the two orthogonal directions $\swarrow\mkern-18mu\nearrow$
and $\searrow\mkern-18mu\nwarrow$ conveyed
in the two output ports and detected thereafter, respectively.

The process can be formalized as follows.
For a two-state process, a two-dimension Hilbert space suffices.
The role of the beam splitter can be described by a very general unitary transformation
which
can be represented by the product of a $U(1)$ phase $e^{-i\,\beta}$ and
of a unimodular unitary
matrix $SU(2)$~\cite{murnaghan}
\begin{equation}
\textsf{\textbf{ T}} (\omega ,\alpha ,\varphi )=
\left(
\begin{array}{cc}
{e^{i\,\alpha }}\,\cos \omega
&
{-e^{-i\,\varphi }}\,\sin \omega
\\
{e^{i\,\varphi }}\,\sin \omega
&
{e^{-i\,\alpha }}\,\cos \omega
 \end{array}
\right)
 \quad ,
\label{2009-e-QvPRSU2}
\end{equation}
where $-\pi \le \beta ,\omega \le \pi$,
$-\, {\pi \over 2} \le  \alpha ,\varphi \le {\pi \over 2}$.
For our purpose, it suffices to consider a 50:50 beam splitter~\cite{Mandel-Ou1987118,green-horn-zei,zeilinger:882,svozil-2004-analog}
of the Hadamard form
$
\textsf{\textbf{H}}= {1\over \sqrt{2}}
\left(
\begin{array}{rr}
1&1\\
1&-1
\end{array}
\right)$, which can be obtained from the general form by setting  $\omega =\frac{\pi}{4}$  and
$\alpha = \beta = \gamma =-\frac{\pi}{2}$  in $e^{-i\,\beta}$ and in Eq.~(\ref{2009-e-QvPRSU2}).
Note that $\textsf{\textbf{H}} \cdot \textsf{\textbf{H}} = \mathbb{I}_2$ is just the identity matrix in two dimensions.

If
$\vert \swarrow\mkern-18mu\nearrow \rangle \equiv (1,0)^T$ and
$\vert \searrow\mkern-18mu\nwarrow \rangle \equiv (0,1)^T$
--- alternatively, we could have used the notation $\vert 0 \rangle$  for $\vert \swarrow\mkern-18mu\nearrow \rangle$,
and $\vert 1 \rangle$  for $\vert \searrow\mkern-18mu\nwarrow \rangle$ ---
represent
certain orthogonal (linear polarization) states measured,
and the particle has been prepared for in a (linear polarization) state
\begin{equation}
\vert \updownarrow \rangle  =
\textsf{\textbf{H}} \vert \swarrow\mkern-18mu\nearrow \rangle =
\frac{1}{\sqrt{2}} \left( \vert \swarrow\mkern-18mu\nearrow \rangle
  + \vert \searrow\mkern-18mu\nwarrow  \rangle \right) \equiv \frac{1}{\sqrt{2}} (1,1)^T,
\end{equation}
which is a 50:50 superposition of both of these states,
then the probability to find the particle in either one of the detectors corresponding to
$\vert \swarrow\mkern-18mu\nearrow \rangle $ and
$\vert \searrow\mkern-18mu\nwarrow  \rangle $ is
\begin{equation}
\begin{array}{rcl}
P_\updownarrow (0)
&=&
\textrm{Tr} \left[
\vert \updownarrow \rangle \langle  \updownarrow \vert \cdot \vert \swarrow\mkern-18mu\nearrow \rangle \langle  \swarrow\mkern-18mu\nearrow \vert
\right]
\equiv
\textrm{Tr}
\left[
\frac{1}{2}\left(
\begin{array}{cc}
1&1\\
1&1
\end{array}
\right)
\cdot
\left(
\begin{array}{cc}
1&0\\
0&0
\end{array}
\right)
\right]
=\frac{1}{2},\;\textrm{
and }\\[4ex]
P_\updownarrow (1)&=&
\textrm{Tr} \left[
\vert \updownarrow \rangle \langle  \updownarrow \vert \cdot \vert \searrow\mkern-18mu\nwarrow  \rangle \langle   \searrow\mkern-18mu\nwarrow \vert
\right]
\equiv
\textrm{Tr}
\left[
\frac{1}{2}\left(
\begin{array}{cc}
1&1\\
1&1
\end{array}
\right)
\cdot
\left(
\begin{array}{cc}
0&0\\
0&1
\end{array}
\right)
\right]
=\frac{1}{2},
\end{array}
\end{equation}
that is, one obtains  a 50:50 chance for the occurrence of outcome $0$ or $1$, respectively.

In general it will be very difficult to establish and maintain an exact (anti)alignment of the polarizers,
resulting in a bias towards either state
$\vert \swarrow\mkern-18mu\nearrow \rangle $ or
$\vert \searrow\mkern-18mu\nwarrow  \rangle $.
If and only if this bias is stationary and the events are independent; i.e., uncorrelated,
then the bias can be eliminated after the coding stage
by von Neumann's normalization procedure~\footnote{
``To cite a human example, for simplicity, in tossing a coin it is probably easier to make two consecutive tosses independent than to toss heads with probability exactly one-half.
If independence of successive tosses is assumed, we can reconstruct a 50--50 chance out of even a badly biased coin by tossing twice.
If we get heads-heads or tails-tails, we reject the tosses and try again. If we get heads-tails (or tails-heads),
we accept the result as heads (or tails).''}:
The biased raw sequence of zeroes and ones is partitioned into
fixed subsequences of length two; then the even parity sequences ``$00$'' and ``$11$'' are discarded,
and only the odd parity ones ``$01$'' and ``$10$'' are kept.
In a second step, the remaining sequences could be mapped into the single symbols $01 \mapsto 0$ and  $10 \mapsto 1$,
thereby extracting a new unbiased sequence at the cost of a loss of original bits~\cite[p. 768]{von-neumann1}
(see Refs.~\cite{elias-72,PeresY-1992} for an improvement of this method,
and Refs.~\cite{stipcevic4442,dichtl-2007,Lacharme-2008} for a discussion of other methods).
This method fails if the events are (temporally) correlated and thus not independent. Take, for instance,
the sequences $010101 \cdots $ or $101010 \cdots $, which in the von Neumann scheme get transformed into $000\cdots$ or  $111\cdots$.
Less spectacular failures of the von Neumann normalization can be constructed by considering convex combinations of these cases.

For beam splitters, the independence of outcomes required by the von Neumann normalization translates into
the assumption that there are no temporal correlations.
In view of the Hanbury Brown Twiss effect (cf., Ref.~\cite[p.313]{chau} and Ref.~\cite[p.127 ff]{knight-qo}),
this assumption is highly nontrivial, as effects of photon bunching might disturb the assumption of independence of subsequent ``quantum coin tosses.''
In particular, it seems that the bit rate might affect the long term statistical independence.
Note also that the von Neumann normalization (cf. above) would fail because of the lack of independence~\cite[p. 768]{von-neumann1} .
Indeed, for ``very high'' (with respect to the regime of the Hanbury Brown Twiss effect) data rates, independence can no longer be assumed.

\subsubsection{Ignorance resulting in a mixed state}

A second, maybe faster and technically less demanding possibility to produce quantum random bits
does not require any preparation step, but just {\em assumes}
the input state to be principally unknowable and indeterminate.
In this case, the system is in a non-pure, mixed state,
reflecting our ignorance about the state prepared~\cite[2nd part, \S~10, p.~827]{schrodinger}.

If the particle is in a totally mixed state,
its density matrix is just proportional to the identity matrix
$\rho_{\mathbb{I}_2 }=\frac{1}{2}\left(
\vert 0 \rangle \langle  0 \vert + \vert 1  \rangle \langle  1 \vert
\right) \equiv \frac{1}{2} \textrm{ diag}\left[\left(1,0 \right) +\textrm{ diag}\left(0,1 \right)\right] = \frac{1}{2}\mathbb{I}_2   $,
and thus the probability to find the particle in either one of the detectors corresponding to
$\vert 0 \rangle $ and
$\vert 1  \rangle $ is
\begin{equation}
\begin{array}{rcl}
P_{\rho_{\mathbb{I}_2 }}(0)&=&
\textrm{Tr} \left[
\rho_{\mathbb{I}_2 } \cdot \vert 0 \rangle \langle  0 \vert
\right]
\equiv
\textrm{Tr}
\left[
\frac{1}{2}\mathbb{I}_2
\cdot
\textrm{ diag}\left(1,0 \right)
\right]
=\frac{1}{2},\;\textrm{
and } \\[2ex]
P_{\rho_{\mathbb{I}_2 }}(1)&=&
\textrm{Tr} \left[
\rho_{\mathbb{I}_2 } \cdot \vert 1  \rangle \langle  1 \vert
\right]
\equiv
\textrm{Tr}
\left[
\frac{1}{2}\mathbb{I}_2
\cdot
\textrm{ diag}\left(0,1 \right)
\right]
=\frac{1}{2};
\end{array}
\end{equation}
that is, one again obtains a 50:50 chance for the occurrence of outcome $0$ or $1$, respectively.

Alas, is may be difficult to certify, control and assert ``ontologically objective,'' as compared to ``epistemically subjective,'' ignorance.
Indeed, the experimenter
preparing the system may {\em subjectively} assume to be ignorant,
whereas the system may implicitly be in a pure state with respect to a certain context, of which
the experimenter does not possess any knowledge, nor has any control.
Also temporal correlations may interfere with randomness.

Note also that any beam splitter is essentially a reversible, one-to-one
``translation device'' ``funneling in'' particles in a certain state, thereby transforming
the state and ``spitting out'' the particles in a bijective manner.
This is reflected in the unitarity of its quantum mechanical description
by the product of $e^{-i\,\beta}$ and  Eq.~(\ref{2009-e-QvPRSU2}).
Ideally, the original signal can be reconstructed and recovered by the serial composition
of the original beam splitter and its ``inverse'' beam splitter associated with the inverse unitary transformation.
In this sense
any quantum random number sequence based on beam splitters is as good as the original source of particles, regardless of the successive (quasi-irreversible) measurement by detectors.

For the sake of demonstration, consider a ``black box'' which,
for undisclosed reasons, contains an (unknown) cyclic particle source or,
if one prefers, a mischievous demon  constantly
releasing particles (emanating from the black box) whose states oscillate
between
$\vert 0' \rangle =
\textsf{\textbf{H}} \vert 0 \rangle =
({1}/{\sqrt{2}}) \left( \vert 0 \rangle
  + \vert 1  \rangle \right) \equiv ({1}/{\sqrt{2}}) (1,1)^T$
and
$\vert 1' \rangle =
\textsf{\textbf{H}} \vert 1 \rangle =
({1}/{\sqrt{2}}) \left( \vert 0 \rangle
  - \vert 1  \rangle \right) \equiv ({1}/{\sqrt{2}}) (1,-1)^T
$, with some frequency $\nu$, such that the state as a function of time is
either (pure case)
\begin{equation}
\vert \varphi_\nu (t) \rangle =  \sin (2\pi \nu t)
\vert 0' \rangle
+
 \cos (2\pi \nu t)
\vert 1' \rangle ,
\end{equation}
or (mixed case)
\begin{equation}
\rho_\nu (t)  =  \sin (2\pi \nu t)
\vert 0' \rangle \langle 0'  \vert
+
 \cos (2\pi \nu t)
\vert 1' \rangle \langle 1'  \vert
.
\end{equation}
If the sampling frequency (or any integer multiple thereof)
of this ``random'' sequence does not coincide with the oscillation frequency $\nu$,
then it may be very difficult for an experimenter to determine the source's regular behavior,
which --- through the beam splitter ---  translates one-to-one into the sequence generated,
since
$\textsf{\textbf{H}}
\vert 0' \rangle
= \textsf{\textbf{H}} \cdot \textsf{\textbf{H}}  \vert 0 \rangle =
 \vert 0 \rangle
$
and
$\textsf{\textbf{H}}
\vert 1' \rangle
=        \textsf{\textbf{H}} \cdot \textsf{\textbf{H}}  \vert 1 \rangle =
\vert 1 \rangle
$.

Thus, it is not totally unjustified to state that claims of ``objective'' randomness
have to be cautiously reviewed
when particles emanating from an underspecified source are targeted directly towards some beam splitter,
as seems to be the case in one of the two setups in Ref.~\cite[Fig.~1(a)]{zeilinger:qct}
and for other devices\footnote{In its {\em White Paper on Random Numbers Generation
using Quantum Physics}~\cite{Quantis}, {\em id Quantique} on p.~7 (in the caption to Fig.~1) announces that
its {\em Quantis} device  uses a light emitting diode,
while at the same time (top of p.~7) pointing out that the monitoring of a Zener diode is problematic:
``Formally the evolution of these generators is not random, but just very complex. One
could say that determinism is hidden behind complexity.''
}.
The quality of the quantum random
sequences produced thus seems to depend on the quality of the light source~\cite{stefanov-2000}
in combination with the beam splitter.
While {\em ``for all practical purposes}''
it may be justified to use a particular (or maybe even any type of) particle source
in combination with a particular beam splitter, this
falls short of a certified procedure to obtain truly random bits
in accordance with Bohm's principle of indeterminacy.

\subsection{Complementary contexts}

Complementarity is a quantum resource for randomness
which may be supporting the random occurrence of individual events
dealing with a mismatch between state preparation and measurement,
as has already been discussed in the Section~\ref{2009-QvPR-s-mismatch}.
It is, however, no sufficient criterion for indeterminism, as can be seen from finite automata~\cite{e-f-moore} or generalized
urn models~\cite{wright}, which are nondistributive but still allow a classical representation~\cite{svozil-2001-eua,svozil-2005-ln1e}.
Whether or not complementarity is a necessary criterion for quantum indeterminism seems to be debatable.
For the lack of necessity, it may suffice to refer to some recording of individual outcomes of
``irreversible'' measurements associated with a ``state reduction,''
or to some decay of a meta-stable state.
Yet, in the first ``state reduction'' case, the existence of principally unpredictable outcomes
seems to be linked to complementarity; at least from an operational point of view.
And also decays of excited states, due to the quantum Zeno effect~\cite{misra:756},
depend on the mode of their measurement, which may be linked to time and energy.
We shall not discuss these issues related to necessity further.

Early discussions of complimentary-type features of quantum mechanics \cite{Heisenberg-27,vonNeumann:1927:WAQ}
concentrate on a finite form of paradoxical self-reference among complementary observables resembling recursion theoretic diagonalization. In the words of Dirac~\cite[\S 1]{dirac},
\begin{quote}
{  ``It is usually assumed that, by being careful, we may cut down the
disturbance accompanying our observation to any desired extent.
The concepts of big and small are then purely relative and refer to the
gentleness of our means of observation as well as to the object being
described. In order to give an absolute meaning to size, such as is
required for any theory of the ultimate structure of matter, we have
to assume that there is a limit to the fineness of our powers of observation
and the smallness of the accompanying disturbance---a limit which is
inherent in the nature cf things and can never be surpassed by improved
technique or increased skill on the part of the observer. If the object under
observation is such that the unavoidable limiting disturbance is negligible,
then the object is big in the absolute sense and we may apply
classical mechanics to it. If, on the other hand, the limiting disturbance
is not negligible, then the object is small in the absolute
sense and we require a new theory for dealing with it.

A consequence of the preceding discussion is that we must revise
our ideas of causality. Causality applies only to a system which is
left undisturbed. If a system is small, we cannot observe it without
producing a serious disturbance and hence we cannot expect to find
any causal connexion between the results of our observations.
Causality will still be assumed to apply to undisturbed systems and
the equations which will be set up to describe an undisturbed system
will be differential equations expressing a causal connexion between
conditions at one time and conditions at a later time. These equations
will be in close correspondence with the equations of classical
mechanics, but they will be connected only indirectly with the results
of observations. There is an unavoidable indeterminacy in the calculation
of observational results, the theory enabling us to calculate in
general only the probability of our obtaining a particular result when
we make an observation.''}
\end{quote}

In 1933, Pauli gave the first explicit definition of complementarity stating that (cf.~\cite[p.~7]{pauli:58},
partial English translation in Ref.~\cite[p.~369]{jammer:89})~\footnote{
{  ``Bei der Unbestimmtheit einer Eigenschaft eines Systems bei einer bestimmten Anordnung
(bei einem bestimmten Zustand eines Systems) vernichtet jeder Versuch, die betreffende Eigenschaft zu messen,
(mindestens teilweise) den Einflu\ss~
der fr{\"u}heren Kenntnisse vom System auf die (eventuell statistischen) Aussagen
{\"u}ber sp{\"a}tere m{\"o}gliche Messungsergebnisse.
[[$\ldots$]]
So m{\"u}ssen, um den Ort eines Teilchens zu bestimmen und um seinen Impuls zu bestimmen,
{\em einander ausschlie\ss ende Versuchsanordnungen benutzt werden.}
[[$\ldots$]]
Die Beeinflussung des Systems durch den Messaparat f{\"u}r den Impuls (Ort)
ist eine solche, da\ss~ innerhalb der durch die Ungenauigkeitsrelationen gegebenen Grenzen
die Benutzbarkeit der fr{\"u}heren Orts- (Impuls-)
Kenntnis f{\"u}r die Voraussagbarkeit der Ergebnisse sp{\"a}terer Orts- (Impuls-) Messungen verlorengegangen ist.
Wenn aus diesem Grunde die Benutzbarkeit {\em eines} klassischen Begriffes in einem
ausschlie\ss enden Verh{\"a}ltnis zu einem {\em anderen} steht, nennen wir diese beiden Begriffe (z.B. Orts- und
Impulskoordinaten eines Teilchens) mit Bohr {\em komplement{\"a}r.}
[[$\ldots$]]
Man wird sehen, dass diese ``Komplementarit{\"a}t'' kein Analogon in
der klassischen Gastheorie besitzt, die ja auch mit statistischen
Gesetzm\"a\ss igkeiten operiert.
Diese Theorie enth{\"a}lt n{\"a}mlich nicht die erst durch die Endlichkeit des Wirkungsquantums
geltend werdende Aussage, da\ss~ durch Messungen an einem System die durch fr{\"u}here Messungen gewonnenen Kenntnisse
{\"u}ber das System unter Umst{\"a}nden notwendig verlorengehen m{\"u}ssen, d.h. nicht mehr verwertet werden k{\"o}nnen.''
}
},
\begin{quote}
{  ``In the case of  an indeterminacy of a property of a system at a certain configuration
(at a certain state of a system), any attempt to measure the respective property (at least partially)
annihilates the influence of the previous knowledge of system on the (possibly statistical) propositions
about possible later measurement results.
[[$\ldots$]]
The impact
on the system by the  measurement apparatus for momentum (position) is such that
within the limits of the uncertainty relations
the value of the knowledge of the previous position (momentum) for the
prediction of later measurements of position and momentum is lost.
If, for this reason, the applicability of {\em one} classical concept stands in the relation of
exclusion to that of {\em another}, we call both of these
concepts (e.g., the position and momentum coordinates of a particle) with Bohr {\em complementary.}
[[$\ldots$]]
One will see that this ``complementarity''
has no analogy in the classical statistical theory of gases,
which also operates with statistical laws.
This theory does not contain the assertion --- which is only valid through the finiteness of the
quantum of action --- that the measurement of a system may necessarily result in a loss
of knowledge acquired through previous measurements; i.e., the previous
measurements can no longer be used.''
}
\end{quote}

Complementarity may thus be interpreted as a subtle kind of departure from
classical omniscience:
whereas it may in principle be possible to measure any single, individual context,
or any (classically operational) observable within (or encompassing) a context,
the direct measurement
(not involving counterfactuals in Einstein-Podolsky-Rosen type configurations~\cite{epr,svozil-2006-omni})
of two or more contexts, or of one context and some observable ``outside'' of it
is impossible.

Until the theorems by Bell, Kochen \& Specker and Greenberger, Horne \& Zeilinger, quantum indeterminism was thus either
(i) ``believed'' and corroborated by the ``effective inability to disprove the contrary'' (i.e., determinism), or
(ii) argued by ``intrinsic self-reference'' and the impossibility of the measurement process
 to act ``softer than'' the quantum of action $h$ on the object.
In the latter case, one could still believe that, contrary to (i), there exist {\em elements of physical reality}, which,
in the sense of Einstein, Podolsky and Rosen~\cite{epr} could even be  measured and
counterfactually~\cite{svozil-2006-omni} inferred simultaneously~\cite{svozil-2006-uniquenessprinciple}.

\subsection{Value indefiniteness}

In deriving the quantum probabilities
---
which have originally been postulated by Born's rule as an axiom of quantum mechanics
---
from a buildup of classical probabilities within contexts in Hilbert spaces of dimension greater than two,
Gleason's theorem~\cite{Gleason,pitowsky:218,rich-bridge,r:dvur-93}
has motivated many authors to derive
nonlocal~\cite{bell,peres222,hey-red,ghz,mermin-93,zeilinger-epr-98}
as well as local~\cite{specker-60,kochen1,ZirlSchl-65,Alda,Alda2,kamber64,kamber65,peres-91,svozil-tkadlec,cabello-96,cabello:210401}
constraints on the existence of {\em global} truth functions (two-valued measures)
on the {\em entire domain} of quantum observables.
Bell's theorem already statistically indicated the
impossibility of co-existence of certain observables
``exceeding'' a single context, e.g.,
by considering the statistics of listing of possible measurement outcomes
and comparing them to the quantum expectations
\cite{peres222}; and the
Kochen-Specker theorem presented a finite proof (by contradiction)
of the impossibility of their co-existence.

When it comes to interpreting and understanding these results, one difficulty  is a fact
already encountered in the study of complementarity:
whereas the {\em totality} of contexts is not co-measurable,
any {\em individual} context is measurable.
In this sense,
the Kochen-Specker and related~\cite{ghz,mermin-93}
theorems  can be viewed to strengthen complementarity:  not only
is it {\em operationally} impossible to directly~\cite{svozil-2006-uniquenessprinciple} measure
more than a single context (despite counterfactual measurements of two contexts
in Einstein-Podolsky-Rosen type configurations~\cite{epr,svozil-2006-omni}) ---
it is provable impossible to consistently assume any co-existence of all quantum observables
which could in principle be measured~\cite{peres222}. We shall refer to this as {\em value indefiniteness.}

Of course, there are ways to ``cope'' with these findings quasi-classically (quasi-realistically)
the most popular being the ``contextuality''
assumption, which was first put forward by Bell
in an attempt to save a kind of realism~\cite{bohr-1949,bell-66,hey-red,redhead}.
It maintains the physical existence of all conceivable potential observables but assumes that the~\cite{bell-66}
``$\ldots$
result of an observation may reasonably depend
not only on the state of the system  $\ldots$
but also on the complete disposition  of the apparatus,''
which could mean that the outcome of a measurement may depend on its context~\footnote{Other schemes to circumvent the quantum value indefiniteness are through
probabilities defined via paradoxical set decompositions~\cite{pitowsky-82,pitowsky-83}
or by considering certain dense subsets of scarcely interlinked quantum contexts~\cite{meyer:99}.}.

Note that, due to the Born rule --- derivable by Gleason's theorem~\cite{Gleason,r:dvur-93,pitowsky:218} for three- and higher-dimensional Hilbert space ---
the quantum mechanical expectation value $\langle E\rangle_\rho =\textrm{Tr} \left(\rho E\right)$ of an observable corresponding to a hermitean operator $E$
and a physical state $\rho$
does not depend on the context; in particular,  the expectation value $\langle E\rangle_\rho$  of a proposition corresponding to a projector $E$
is independent of the particular choice of basis among the continuity of orthogonal bases which it may belong to~\cite{svozil:040102}.
Thus, contextuality is restricted to {\em single, individual outcomes} of potential measurements.
Stated differently, quantum mechanics does not determine a specific measurement outcome of an observable,
but  determines the expectation value of that observable.
In this respect, the quantum contextuality assumption is somewhat similar to Born's concept of deterministic evolution of the quantum state as compared to the
indeterministic occurrence of single events; or the {\em outcome dependence {\it versus} parameter independence}
for remote nonlocal~\cite{wjswz-98} correlated quantum events~\cite{shimony2}.

The Kochen-Specker theorem is a rather strong indication of value indefiniteness
and thus  of quantum indeterminism~\cite{2008-cal-svo} and randomness beyond Born's conjecture of the random occurrence of
individual events, and even beyond complementarity; at least for multi-context configurations where Kochen \& Specker constructions are viable.

Since a nontrivial interconnectedness of different bases
is possible only for Hilbert spaces of dimension three onwards,
the Gleason and the Kochen-Specker theorems apply only to Hilbert spaces
of dimensions {\em higher than two} (see the related argument in Ref.~\cite[p.~193]{peres}); hence value indefiniteness can be proven
only for systems of {\em three or more mutually exclusive outcomes}.
For two-dimensional systems, one has still to rely purely on Born's indeterminacy postulate,
solely backed by complementarity and the quantum uncertainty relations.
We have to conclude that,
as presently many  quantum random number generators using beam splitters (also the ones utilizing complementarity)
operate with two exclusive outcomes,
they are not backed by value indefiniteness in the sense of Bell, Kochen \& Specker and Greenberger, Horne \& Zeilinger.

One may still argue that, although the Born rule for quantum probabilities and expectations
cannot be proven from the (more elementary) assumptions of Gleason's theorem~\cite[\S~7.2]{peres}
for two-dimensional Hilbert spaces by presently known mathematical methods, this
does not exclude the possibility that some other methods exist which would  prove similar results
related to value indefiniteness even for physical configurations with two
mutually exclusive outcomes.
For the sake of excluding this latter possibility, one should,
for instance, find a counterexample
(on the structure of quantum observables in two-dimensional Hilbert space)
which
(i) either is not in accordance with the Born rule but still
in accordance with the additivity property upon which Gleasons's theorem is based;
(ii) or is in accordance with the Born rule but allows
two-valued states which may or may not be sufficient for a homeomorphic embedding into a Boolean algebra.
A typical counterexample of the first type
would be one in which an electron spin observable,
for noncollinear directions, would always point ``up'' and ``down'' according to some algorithmic
rule~\cite[pp.~70-72]{svozil-ql}).
Formally, this is due to the fact that, for two-dimensional configurations, there exists a
full, separating set of two-valued states.
A counterexample of the second type appears to allow merely states which
are singular only in a {\em single} pair of observables (indeed, this is true for arbitrary Hilbert space dimensions), and thus are insufficient for the particular purpose.

\subsection{Incomputability of quantum randomness and empirical testing}
\label{incomput}

In~\cite{2008-cal-svo} it is proved that quantum randomness is not Turing computable.
More precisely, suppose that a quantum experiment produces an infinite sequence of quantum random bits.
Would such a sequence be computable by a Turing machine?
If we accept value indefiniteness as expressed by the theorems of Bell, Kochen \& Specker and Greenberger, Horne \& Zeilinger,
then the answer given in Ref.~\cite{2008-cal-svo} is negative;
even more, {\it no Turing machine can enumerate an infinity of correct bits of such a sequence}.
For example, an infinite sequence of quantum random bits may start with a billion of 0's,
but cannot consist entirely of only 0's. The infinite sequence of  bits $0100011011000001\ldots$
(Champernowne's constant) or the binary expansion of $\pi$ cannot be exactly reproduced by any quantum experiment.

But is quantum randomness a ``true'' and ``objective'' form of randomness?
First, and foremost, there is no such thing as ``true'' randomness as measure-theoretical arguments show~\cite{calude:02}.
Secondly, {\it  it is an open question  whether quantum randomness satisfies the
requirements of algorithmic randomness}~\cite{calude:02}.

Our aim is to experimentally study the possibility of distinguishing between
quantum sources of randomness (proved to be theoretically incomputable) and  some well-known computable
sources of  pseudo-randomness.
 The  legitimacy of the experimental approach comes from algorithmic information theory
which provides characterizations of algorithmic random sequences in terms of the degrees of incompressibility of their finite prefixes.
More precisely, a sequence is algorithmic random iff  all its finite prefixes cannot be compressed by a universal prefix-free Turing machine by more than a fixed constant
(which depends on the fixed machine and sequence and not on prefixes)~\cite{calude:02}.
The degree of incompressibility of a string is measured with the prefix-complexity $H_{U}$ (which depends on the universal prefix-free Turing machine
$U$). The best empirical test of randomness would be to calculate the prefix-complexity of all prefixes of  a given (long) string.
This is impossible because the prefix-complexity is incomputable.
However, there are computable, but weaker properties than incompressibility which can be tested on prefixes, for example, Borel normality (explained below).
Of course, any such property is necessary, but not sufficient; hence  the (degree of) {\it absence of the property is significant.}

We have performed
tests of randomness on pseudo-random strings (finite sequences) of length $2^{32}$ generated with software
(Mathematica, Maple),  which are not only computable, but also cyclic, the bits of $\pi$, which is computable, but not cyclic, and strings produced by quantum measurements with the commercial device Quantis, as well as  by the Vienna IQOQI group.

The signals of the Vienna Institute for Quantum Optics and Quantum
Information (IQOQI)
group were generated with photons from a weak blue LED light source which impinged on a
beam splitter
without any polarization sensitivity with two output ports associated
with the codes ``0'' and ``1,'' respectively~\cite{zeilinger:qct}.
There was no pre- or post-processing of the raw data stream,
however the output was constantly monitored
(the exact method is subject to a patent pending).
In very general terms, the setup needs to be running for at least one
day to reach a stable operation.
There is a regulation mechanism which keeps track of the bias between
``0'' and ``1,''
and  tunes the random generator for perfect symmetry.
Each data file was created in one continuous run of the device
lasting over hours.

Our empirical tests indicate quantitative differences between computable
and incomputable sources by examining (long, but)
finite prefixes of infinite sequences.
Such differences are guaranteed to exist by the result in Ref~\cite{2008-cal-svo}, but, because computability is an asymptotic property,
there is no guarantee that
finite tests can ``pick'' them  in the prefixes we have analyzed.
We performed more tests than those described below, but discarded
those for which the results were inconclusive  (cf. Ref.~\cite{Rukhin-nist}).
In what follows we will describe
a battery of ``non-standard'' randomness tests based on coding theory and
algorithmic information theory results~\cite{calude:02}  which distinguish
between the  computable
and incomputable sources that we sampled.

\section{Randomness tests}

\label{tests}
In order to avoid some ideological or metaphysical bias, all
sequences have been treated on an equal footing by looking with
``evenly-suspended attention'' at their phenomenological encoded
phenotypes. No hidden ``meaning'' or ``message'' should be ascribed to them.
This is conceptually related to the following scenario.

Consider a couple of labeled ``black boxes,'' each being the source
of binary sequences, emanated at a constant rate.
In our case, we have two ``Born boxes'' operating under Born's
assumption of quantum randomness (actually, Quantis is just that),
a ``Pi box'' humming out binary digits of $\pi$, as well as some
``Sinners'' (in von Neumann's judgment~\cite{von-neumann1})
containing algorithms pretending to
output random digits.

Suppose that these boxes cannot be ``screwed open,'' and no clues
about the origin of the symbolic sources are otherwise obtainable
from the outside in any perceivable way.
Suppose further that somebody (either a devil, or a malign colleague,
or a cleaning agent) has erased the labels completely.
Would we be able to tell which box is which by analyzing their bit
renditions alone?
In what follows, we shall present some tentative answers to this question
based on data produced with these boxes.

 \subsection{Data}
Our data consist of 50 binary sample ``random'' strings  of length
$2^{32}$: 10 pseudo-random strings  produced by Mathematica 6~\cite{MRG},
10 pseudo-random strings  produced by  Maple 11~\cite{MAPLE}, 10 quantum random strings
generated with  Quantis~\cite{Quantis}, 10 quantum random strings
generated by the Vienna IQOQI group~\cite{Vienna},
and  10 strings of $2^{32}$ bits from the binary expansion of $\pi$ obtained from~\cite{pi}.

The process used to generate ten strings from $\pi$ is the following.   The input was given
to us in hexadecimal format, with two decimal digits per byte; or one decimal digit per nibble.
Two random decimal digits were selected to be omitted throughout the string~\footnote{For the curious, our ten pairs
of deleted digits were $\{0,1\},\{0,5\},\{1,6\},\{2,3\},\{2,7\},\{3,8\},\{4,5\},\{4,9\},\{6,7\},$
and $\{8,9\}$.} The remaining decimal digits are assigned a 3-bit binary number 0 to 7,
which are output as 3 bits each.
Processing continues until $2^{32}$  bits are output.
The input source that we downloaded had 4,200,000,000 decimal digits so potentially up to
$1.008\times 10^{10}$ bits can be extracted (which is about $2.347 \times 2^{32}$);
thus
almost all of these digits are needed to generate our 10 strings. The
justification that these ``projected'' binary strings share the same randomness properties of
$\pi$ is given by the following result~\cite{calude:02}: if in a random sequence over an alphabet $\{a_{1}, \ldots , a_{k}\}$, $k>2$, we remove all occurrences of a
fixed symbol $a_{i}$, then the new sequence is also random (over
an alphabet with $k-1$ symbols).

\subsection{Descriptive statistics}

  Our experiments have been uniformly performed on all these fifty sample strings.
The tests presented below can be grouped into the following classes:
\begin{enumerate}
\item
Borel normality test;
\item
test based on Shannon's information theory;
\item
two tests  based on algorithmic information theory; and
\item
test based on random walks.
\end{enumerate}

We present our test results using
box-and-whisker plots which are compact graphical representations of
groups of numerical data through five characteristic summaries:
 test minimum value, first quantile (representing one fourth of the test data), median or second quantile (representing half of the test data), third quantile (representing three fourths of the test data), and test maximum value.
Mean and standard deviation of the data representing the results of the tests are calculated.
For the reader who prefers ``numbers'' instead of ``pictures,'' tables containing
all these seven elements of descriptive statistics are included for all five sources.

Tables containing the experimental data and the programs used
to generate the data can be downloaded from our extended paper~\cite{CDMTCS372}.

\subsubsection{Borel normality test}

Borel normality was the first mathematical definition of randomness~\cite{borel:09}.
A sequence is (Borel) normal if every  binary string appears in the sequence with the right probability
(which is $2^{-n}$ for a string of length $n$).
A  sequence is normal if and only  it  is incompressible by any information lossless finite-state compressor~\cite{ZL:1978}, so  normal sequences are  those sequences  that appear random to any finite-state machine.

Every algorithmic random infinite sequence is Borel normal~\cite{DBLP:conf/dlt/Calude93}.
 The converse implication is not true:
there exist computable normal sequences (e.g.\  Champernowne's constant).

Normality is invariant under finite variations: adding, removing, or changing a finite number of bits in any normal sequence leaves it normal. Further, if a sequence satisfies the normality condition
for strings of length $n+1$, then it also satisfies normality for strings of length $n$, but the converse is not true.

Normality was transposed to  strings in Ref.~\cite{DBLP:conf/dlt/Calude93}. In this process one  has to replace limits with inequalities. As a consequence, the above two properties, which are valid for sequences, are no longer true for strings.

For any fixed  integer $m > 1$, consider the alphabet $B_{m} = \{0,1\}^{m}$ consisting of all binary strings of length $m$,
 and for
every $1 \leq i \leq 2^{m}$ denote by $N_{i}^{m}$ the
 number of occurrences of the lexicographical $i$th binary string of length $m$ in the string $x$ (considered over the alphabet $B_{m}$).
 By $|x|_{m}$ we denote the length of $x$.
 A string $x$ is Borel normal if
  for every natural $1 \leq m \leq \log_{2}\log_{2} |x|,$
  \[
\left| \frac{N_{j}^{m}(x)}{|x|_{m}} -  2^{-m} \right| \leq
\sqrt\frac{\log_{2}|x|}{|x|}\raisebox{0.5ex}{,}\]
for every $1 \leq j \leq 2^{m}$.
In Ref.~\cite{DBLP:conf/dlt/Calude93}
it is shown that almost all algorithmic random strings are Borel normal.

\begin{figure}[htbp] 
   \centering
   \includegraphics[width=6in]{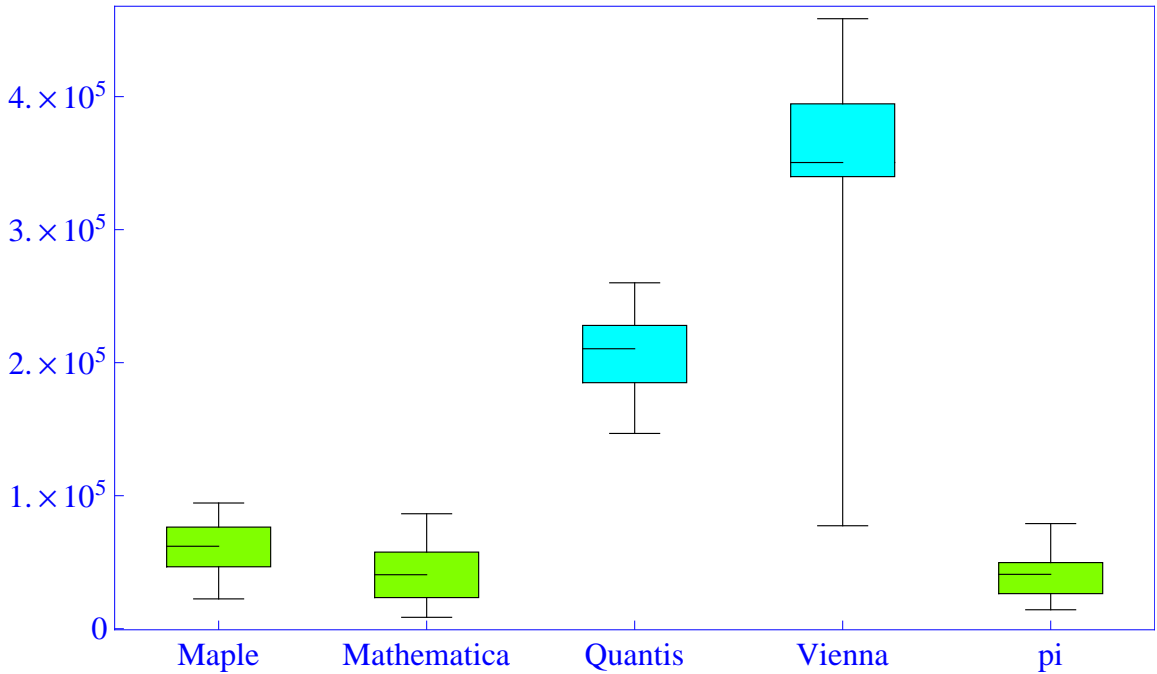}
   \caption{(Color online) Box-and-whisker plot for the results for tests of the Borel normality property.}
   \label{fig:example1}
\end{figure}

In the first test
we  count the maximum, minimum and difference of non-overlapping occurrences of  $m$-bit ($m=1,\ldots , 5$) strings  in
each sample string.   Then we tested the Borel normality property for each sample string and found that
almost all strings pass the  test,  with some notable exceptions.
We found that several of the Vienna sequences failed the expected count range for
$m=2$ and a few of the Vienna sequences were outside the expected range for $m=3$
and $m=4$ (some less then the expected minimum count and some more than the expected
maximum count).  The only other bit sequence that was outside the expected range
count was one of the Mathematica sequences that had a too big of a count for $k=1$.
Figure~\ref{fig:example1} depicts a box-and-whisker plot of the
results. This is followed by statistical (numerical) details
in Table~\ref{tab:1}.

\begin{table}
\caption{Statistics for the results for tests of the Borel normality property.}\label{tab:1}
\begin{center}
\begin{tabular}
[c]{c c c c c c c c c c c c c c c }%
\hline\hline
Descriptive statistics & min & Q1 & median & Q3 & max & mean & sd\\\hline
Maple & 22430 & 47170 & 61990 & 76130 & 94510 & 60210 & 21933.52\\
Mathematica & 8572 & 25500 & 40590 & 55650 & 86430 & 41870 & 23229.77\\
Quantis & 146800 & 185100 & 210500 & 226600 & 260000 & 207200 & 33515.65\\
Vienna & 77410 & 340200 & 350500 & 392500 & 260000 & 337100 & 103354.3\\
$\pi$ & 14260 & 28860 & 40880 & 47860 & 79030 & 40220 & 17906.21\\\hline\hline
\end{tabular}
\end{center}
\end{table}

\subsubsection{Test based on Shannon's information theory}

\begin{figure}[htbp] 
   \centering
   \includegraphics[width=6in]{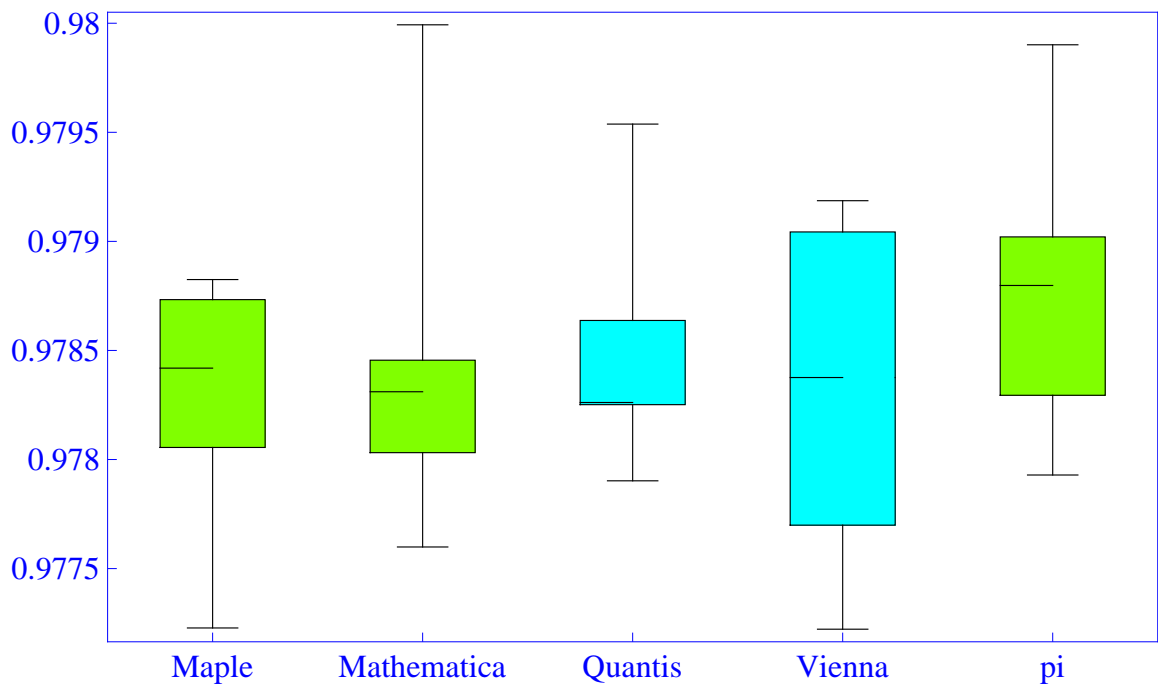}
   \caption{(Color online) Box-and-whisker plot for average results in ``sliding window'' estimations of the Shannon entropy.}
   \label{fig:example2}
\end{figure}

The second test
computes  ``sliding window'' estimations of the Shannon entropy $L_n^1, \ldots ,L_n^t$ according to the method described in
\cite{Wyner}: a smaller  entropy is a symptom of  less randomness.
The results are presented in
Figure~\ref{fig:example2} and Table~\ref{tab:2}.

\begin{table}
\caption{Statistics for average results in ``sliding window''
estimations of the Shannon entropy.}\label{tab:2}
\begin{center}
\begin{tabular}
[c]{c c c c c c c c c c c c c c c }%
\hline\hline
Descriptive statistics & min & Q1 & median & Q3 & max & mean & sd\\\hline
Maple & 0.9772 & 0.9781 & 0.9784 & 0.9787 & 0.9788 & 0.9783 & 0.0005231617\\
Mathematica & 0.9776 & 0.9781 & 0.9783 & 0.9785 & 0.9800 & 0.9783 & 0.0006654936\\
Quantis & 0.9779 & 0.9783 & 0.9783 & 0.9786 & 0.9795 & 0.9784 & 0.0004522699\\
Vienna & 0.9772 & 0.9777 & 0.9784 & 0.9790 & 0.9792 & 0.9783 & 0.0006955834\\
$\pi$ & 0.9779 & 0.9784 & 0.9788 & 0.9790 & 0.9799 & 0.9788 & 0.0006062724\\\hline\hline
\end{tabular}
\end{center}
\end{table}

\subsubsection{Tests based on algorithmic information theory}

\begin{figure}[htbp] 
   \centering
   \includegraphics[width=6in]{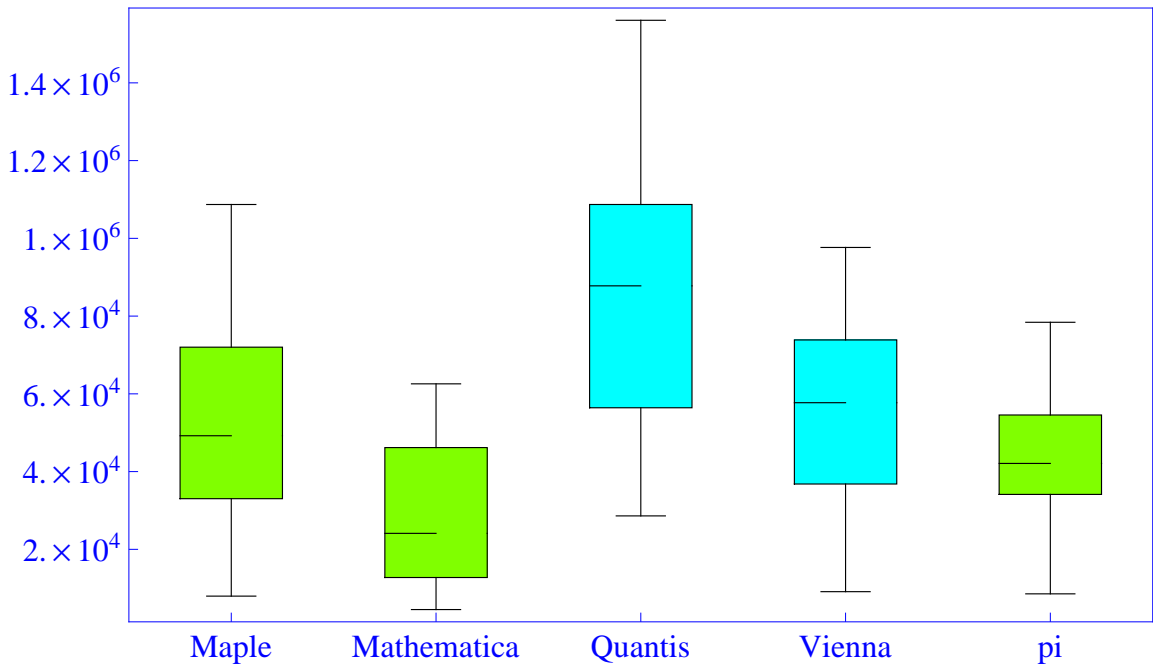}
   \caption{(Color online) Box-and-whisker plot for the results of  the ``book stack'' randomness test.}
   \label{fig:example3}
\end{figure}

The third test
uses  the ``book stack'' (also known as ``move to front'') randomness
test as proposed in Ref.~\cite{MR2099021,MR2162569}.
More compression is a symptom of less randomness.
The results, presented in
Figure~\ref{fig:example3} and Table~\ref{tab:3}, are derived from the original count, the count after the application
of the transformation, and the difference.  The key metric for this test is
the count of ones after the transformation.  The book stack encoder does
not compress data but instead rewrites each byte with its index (from
the top/front) with respect to its input characters being
stacked/moved-to-front.
Thus, if a lot of repetitions occur (i.e., a symptom of
non-randomness), then the output  contains more zeros than ones due to
the sequence of indices generally being smaller numerically.

\begin{table}
\caption{Statistics for the results of  the ``book stack'' randomness
test.}\label{tab:3}
\begin{center}
\begin{tabular}
[c]{ c c c c c c c c }%
\hline\hline
Descriptive statistics & min & Q1 & median & Q3 & max & mean & sd\\\hline
Maple & 7964 & 34490 & 49220 & 69630 & 108700 & 53410 & 33068.58\\
Mathematica & 4508 & 13020 & 24110 & 43450 & 62570 & 27940 & 19406.03\\
Quantis & 28600 & 60480 & 87780 & 106700 & 156100 & 89990 & 41545.76\\
Vienna & 9110 & 38420 & 57720 & 73220 & 97660 & 53860 & 27938.92\\
$\pi$ & 8551 & 35480 & 42100 & 52870 & 78410 & 41280 & 20758.46\\\hline\hline
\end{tabular}
\end{center}
\end{table}

The fourth test is   based solely on the behavior of algorithmic random strings (as
selectors for Solovay-Strassen probabilistic primality test) and not on specific properties of randomness.

To test whether  a positive integer
$n$ is prime, we take $k$ natural numbers uniformly distributed between 1
and $n - 1$, inclusive, and, for each one $i$, check whether the predicate
$W (i, n)$ holds.  If this is the case we say that
``$i$ is a witness of $n$'s compositeness''.
If $W (i, n)$ holds for at least one $i$ then $n$ is
composite; otherwise, the test is inconclusive, but in this case if one declares $n$ to be  prime then the
 probability to be wrong is smaller than $ 2^{-k}$.

 This is due to the fact that
 at least half  $i$'s from $1$ to $ n - 1$ satisfy $W (i, n)$  if $n$ is indeed composite,
 and {\it none} of them satisfy $W (i, n)$ if $n$ is prime~\cite{solovay:84}.
Selecting $k$ natural numbers  between 1
and $n - 1$ is the same as choosing a binary string $s$ of length $n-1$ with $k$ $1$'s
such that the $i$th bit is 1 iff $i$ is selected.
Ref.~\cite{ch-schw-78} contains a  proof that, if $s$ is a long enough   algorithmically random binary string,
then $n$ is prime iff $Z(s,n)$ is true,
where $Z$ is a predicate constructed directly from conjunctions of negations of  $W$~\footnote{
In fact, every  ``decent'' Monte Carlo simulation algorithm in  which  tests are chosen according to an
algorithmic random string produces a result
which is not only true with high probability, but {\it rigorously correct}~\cite{MR757602}.}.

A Carmichael number is a composite positive integer $k$ satisfying the congruence $b^{k-1} \equiv 1 ({\rm mod } \, k)$
for all integers $b$ relative prime to $k$.
Carmichael numbers are composite, but are difficult to factorize and thus are ``very similar'' to primes;
they are sometimes called pseudo-primes.
Carmichael numbers can fool Fermat's primality test, but less   the Solovay-Strassen test.
With increasing values, Carmichael numbers
become ``rare''~\footnote{There are 1,401,644 Carmichael numbers in the interval $[1, 10^{18}]$.}.

\begin{figure}[htbp] 
   \centering
   \includegraphics[width=6in]{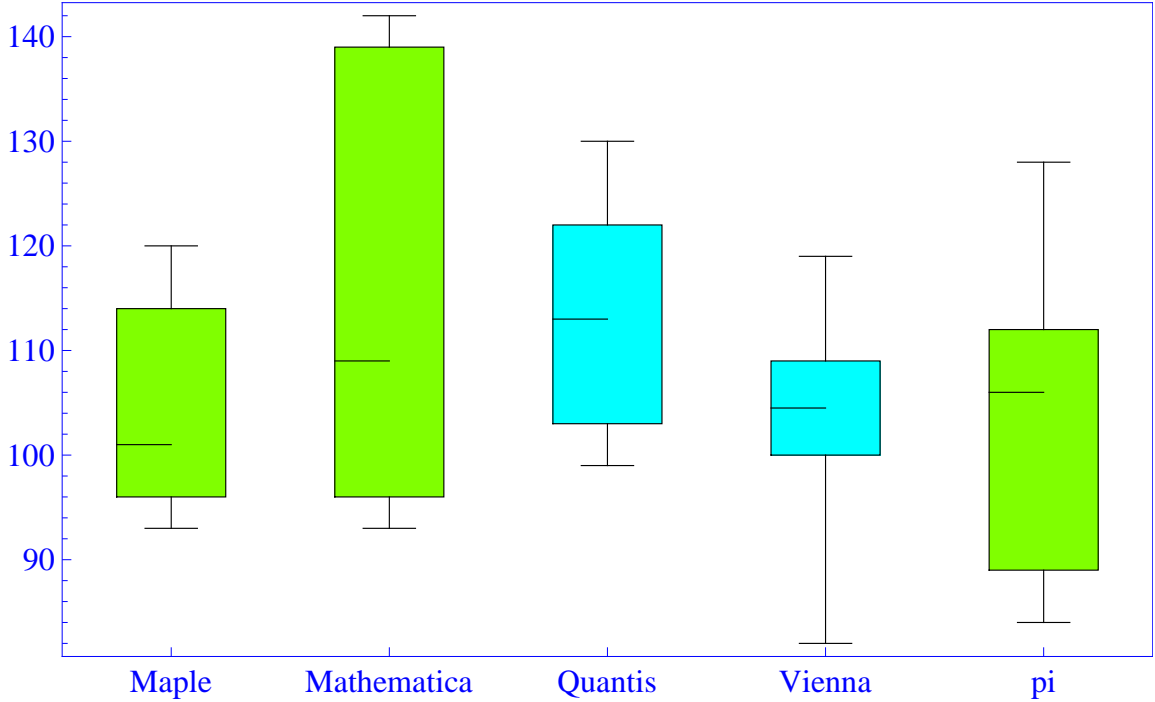}
   \caption{(Color online) Box-and-whisker plot for the results based on the Solovay-Strassen probabilistic primality test.}
   \label{fig:example4}
\end{figure}

The fourth test uses Solovay-Strassen probabilistic primality test for
Carmichael numbers (composite) with prefixes of the sample strings as the binary
string $s$. We used the Solovay-Strassen  test for all Carmichael numbers less
than $10^{16}$---computed in Ref.~\cite{Pinch,Pinch07}---with numbers selected according to increasing prefixes of each sample string till the algorithm returns
a non-primality verdict. The metric is given by the length of the sample used to
reach the correct
verdict of non-primality for all of the 246683 Carmichael numbers less than
$10^{16}$.  [We started with $k=1$ tests (per each Carmichael number) and increase $k$
until the metric goal is met; as $k$ increases we always use new bits (never
recycle) from the sample source strings.]
The results are presented in
Figure~\ref{fig:example4} and Table~\ref{tab:4}.

\begin{table}
   \caption{Statistics for the results based on the
Solovay-Strassen probabilistic primality test.}\label{tab:4}
\begin{center}
\begin{tabular}
[c]{ c c c c c c c c }%
\hline\hline
Descriptive statistics & min & Q1 & median & Q3 & max & mean & sd\\\hline
Maple & 93.0 & 96.0 & 101.0 & 113.5 & 120.0 & 104.9 & 10.57723\\
Mathematica & 93.0 & 97.0 & 109.0 & 132.3 & 142.0 & 113.5 & 19.60867\\
Quantis & 99.0 & 103.3 & 113.0 & 121.3 & 130.0 & 112.6 & 10.66875\\
Vienna & 82.0 & 100.3 & 104.5 & 109.0 & 119.0 & 103.5 & 11.03781\\
$\pi$ & 84.0 & 91.75 & 106.0 & 110.8 & 128.0 & 104.7 & 10.66875\\\hline  \hline
\end{tabular}
\end{center}
\end{table}

\subsubsection{Test based on random walks}


A symptom of non-randomness of a string is detected when the plot generated by viewing a sample sequence as
a 1D random walk meanders less  away
from the starting point (both ways); hence the max-min range is the metric.

\begin{figure}[htbp] 
   \centering
   \includegraphics[width=6in]{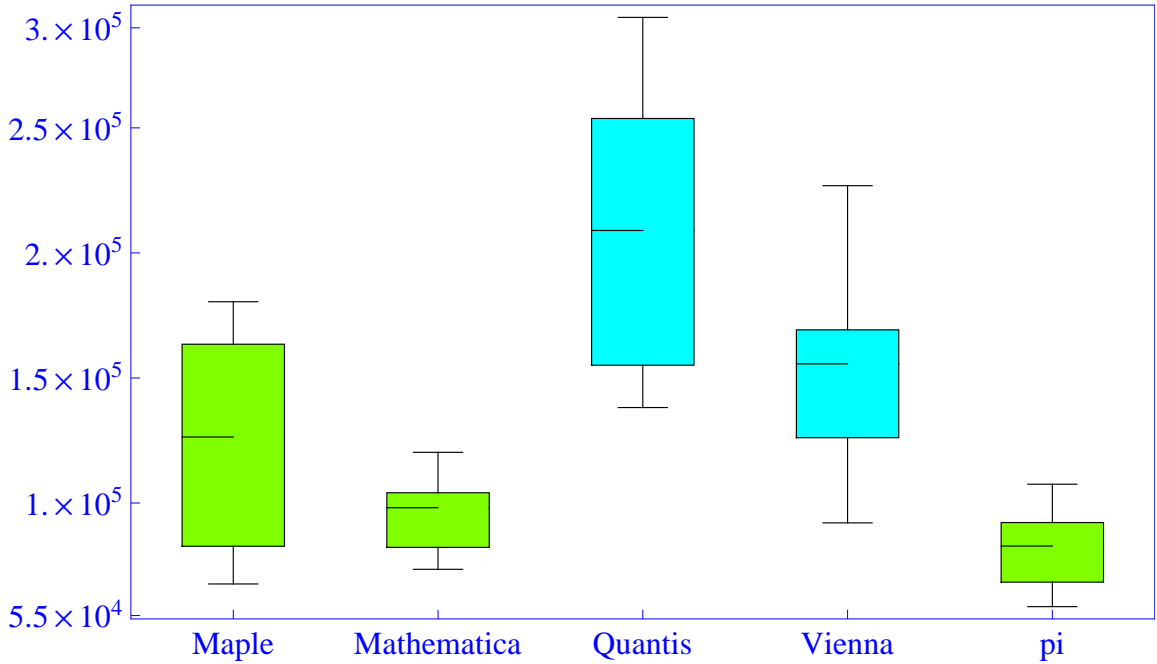}
   \caption{(Color online) Box-and-whisker plot for the results of the random walk tests.}
   \label{fig:example5}
\end{figure}

The fifth test
is based on viewing a  random
sequence as a 1D random walk.  Here the bits (indices along the $x$-axis) are
interpreted as follows: 1=move up, 0=move down ($y$-axis).
This test measures  how far away from the starting point (in either positive
or negative) from the starting $y$-value of 0 that one can reach using
successive bits of the sample sequence.
Figure~\ref{fig:example5} and Table~\ref{tab:5} summarize the results.

\begin{table}
\caption{Statistics for the results of the random walk tests.}\label{tab:5}
\begin{center}
\begin{tabular}
[c]{ c c c c c c c c }%
\hline\hline
Descriptive statistics & min & Q1 & median & Q3 & max & mean & sd\\\hline
Maple & 67640 & 88730 & 126400 & 162500 & 180500 & 125300 & 42995.59\\
Mathematica & 73500 & 84760 & 98110 & 103400 & 120300 & 96450 & 14685.34\\
Quantis  & 138200 & 161600 & 209000 & 250200 & 294200 & 211300 & 55960.23\\
Vienna & 92070 & 130200 & 155600 & 167600 & 226900 & 152900 & 36717.55\\
$\pi$ & 58570 & 70420 & 82800 & 91920 & 107500 & 82120 & 14833.75\\\hline\hline
\end{tabular}
\end{center}
\end{table}

\subsection{Statistical analysis of randomness tests results}

In what follows statistical tests are used to compare the probability
distributions of results of randomness tests applied to the strings generated by the
five sources.
The Kolmogorov-Smirnov test  for two samples \cite{Conover}
tries to determine if two datasets differ significantly.
This test
has the advantage of making no assumption about the distribution of data; i.e., it
is non-parametric and distribution free.
The Kolmogorov-Smirnov test returns a $p$-value,
and the decision ``the difference between the two datasets is statistically
significant" is accepted if the $p$-value {\em is less than $0.05$;}
or, stated pointedly, if the
probability of taking a wrong decision is less than $0.05$. Exact $p$-values are
only available for the two-sided two-sample tests with no ties.

In some cases we have tried to double-check the decision  ``no significant
differences between the datasets"  at the price of a supplementary, plausible
distribution assumption. Therefore, we have performed the Shapiro-Wilk test  for normality
\cite{Shapiro-Wilk} and, if normality is not rejected, we have assumed that the
datasets have normal (Gaussian) distributions.
In order to be able to compare the expected values (means) of the two samples,
the Welch $t$-test~\cite{Welch}, which is a version of Student's test, has been applied.

The Shapiro-Wilk test axamines the null hypothesis that a sample
$z_{1},\ldots ,z_{n}$ comes from a normally distributed population. This test is
appropriate for small samples, since it is not an asymptotic test.
As for each
source ten independent strings have been studied, we have applied the
Shapiro-Wilk test for a sample size $n = 10$.

The Welch's $t$-test~\cite{Welch} is an adaptation of Student's $t$-test  used
with two samples having possibly unequal variances. It is used to test the
null hypothesis that the two population means are equal (using a two-tailed test).

The calculations
have been performed with the software {\em ``R''}~\cite{rproject}.
In order to emphasize the relevance of p-values less than 0.05 associated with Kolmogorov-Smirnov,
Shapiro-Wilk and Welch's $t$-tests, they are printed in boldface and discussed in the text.

\subsubsection{Borel test of normality}

\if01
\begin{center}
\begin{tabular}
[c]{ c c c c c c }\hline\hline
Shapiro-Wilk test & Maple & Mathematica & Quantis & Vienna  & $\pi$\\\hline
$p$-value 
& 0.9183 & 0.966 & 0.9708 &
\bf{0.01474} & 0.4898\\\hline\hline
\end{tabular}
\end{center}

\fi

The results of the Kolmogorov-Smirnov test are presented in Table~\ref{tab:5b}.

\begin{table}
\caption{Kolmogorov-Smirnov test for the Borel normality tests.}\label{tab:5b}
 \begin{center}
 \begin{tabular}
[c]{ c c c c c }
\hline\hline
Kolmogorov-Smirnov test $p$-values & Mathematica & Quantis & Vienna & $\pi$ \\\hline
Maple & 0.4175 &$\mathbf{< 10^{-4}}$
& \bf{0.0002}
& 0.1678\\
Mathematica &  & $\mathbf{< 10^{-4}}$
& \bf{0.0002} & 0.9945\\
Quantis &  &  & \bf{0.0002} & $\mathbf{< 10^{-4}}$
\\
Vienna &  &  &  & \bf{0.0002}\\\hline\hline
\end{tabular}
\end{center}
\end{table}

Statistically significant differences are identified for
(i) Quantis {\it versus}  Maple,  Maple, Mathematica and $\pi$;
(ii) Vienna {\it versus}  Maple, Mathematica and $\pi$; and
(iii) Quantis {\it versus}  Vienna.

Note that

\begin{enumerate}

\item Pseudorandom strings pass the Borel normality test for comparable numbers of counts, relatively small:
if the angle brackets   $\langle x \rangle$ stand for the statistical mean of tests on $x$, then
 $\langle \text{Maple} \rangle  = 60210$, $\langle \text{Mathematica} \rangle  = 41870$,
$\langle  \pi \rangle  = 40220$).

\item Quantum strings pass the Borel normality test only for ``much larger numbers''
of counts
($\langle \text{Quantis} \rangle  = 207200$, $\langle \text{Vienna} \rangle  = 337100$),



\end{enumerate}

As a result, the Borel normality test detects and identifies
statistically significantly differences between all pairs of computable and incomputable  sources of ``randomness.''

 \subsubsection{Test based on Shannon's information theory}

 The results of the Kolmogorov-Smirnov test are presented in  Table~\ref{tab:6}.
No significant differences are detected. The descriptive statistics
data for the results of this test  indicates almost identical
distributions corresponding to the five sources.

\subsubsection{Tests based on algorithmic information theory}

\begin{table}
\caption{Kolmogorov-Smirnov test for Shannon's information theory tests.}\label{tab:6}
 \begin{center}
 \begin{tabular}
[c]{ c c c c c }
\hline\hline
Kolmogorov-Smirnov test $p$-values & Mathematica & Quantis & Vienna & $\pi$ \\\hline
Maple & 0.7870 & 0.7870 & 0.7870 & 0.1678\\
Mathematica &  & 0.7870 & 0.4175 & 0.0525\\
Quantis &  &  & 0.4175 & 0.1678\\
Vienna &  &  &  & 0.4175\\\hline\hline
\end{tabular}
\end{center}
\end{table}

The results of the Shapiro-Wilk test are presented in Table~\ref{tab:7}.
Since there is no clear
pattern of normality for the data, the application of Welch's $t$-test is not
appropriate.

\begin{table}
\caption{Shapiro-Wilk test for Shannon's information theory tests.}\label{tab:7}
\begin{center}
\begin{tabular}
[c]{ c c c c c c }\hline\hline
Shapiro-Wilk test & Maple & Mathematica & Quantis  & Vienna  & $\pi$\\\hline
$p$-value & 0.1962 & \bf{0.0189} &
\bf{0.0345} & 0.3790 & 0.8774\\\hline\hline
\end{tabular}
\end{center}
\end{table}

\if01
Comments:

\begin{itemize}
\item A smaller entropy is a symptom of less randomness, but no significant differences among the mean entropy or its quantiles are identified.

\item 	Normality  was rejected by the Shapiro-Wilk test for Mathematica and and Quantis.

\end{itemize}
\fi

The results of the Kolmogorov-Smirnov test associated with the ``book-stack" tests are enumerated in Table~\ref{tab:8}.
Statistically significant differences are identified for
Quantis {\it versus} Mathematica and $\pi$.

\begin{table}
\caption{Kolmogorov-Smirnov test for the ``book-stack'' tests.}\label{tab:8}
 \begin{center}
 \begin{tabular}
[c]{ c c c c c }
\hline\hline
Kolmogorov-Smirnov test $p$-values & Mathematica & Quantis & Vienna & $\pi$ \\\hline
Maple  & 0.4175 & 0.1678 & 0.9945 & 0.4175\\
Mathematica &  & \bf{0.0021} & 0.1678 & 0.4175\\
Quantis &  &  & 0.1678 & \bf{0.0123}\\
Vienna &  &  &  & 0.4175\\\hline\hline
\end{tabular}
\end{center}
\end{table}

As more compression is a symptom of less randomness,  the corresponding ranking of
samples is as follows:
$\langle \text{Quantis} \rangle  = 89988.9 > \langle \text{Vienna} \rangle  = 53863.8  >  \langle \text{Maple} \rangle  =
53411.6 > \langle  \pi \rangle  = 41277.5  >  \langle \text{Mathematica} \rangle  = 27938.3$.

The Shapiro-Wilk tests results are presented in  Table~\ref{tab:9}.

\begin{table}
\caption{Shapiro-Wilk test for the ``book-stack'' tests.}\label{tab:9}
\begin{center}
\begin{tabular}
[c]{ c c c c c c }\hline\hline
Shapiro-Wilk test & Maple & Mathematica & Quantis & Vienna  & $\pi$\\\hline
$p$-value  & 0.7880 & 0.4819 & 0.7239 & 0.8146 &
0.5172\\\hline\hline
\end{tabular}
\end{center}
\end{table}

Since normality is not rejected for any string, we apply the Welch's $t$-test
for the comparison of means. The results are enumerated in Table~\ref{tab:10}.
Significant differences between the means are identified for the following sources:
(i) Quantis {\it versus} all other sources (Maple, Mathematica, Vienna, $\pi$); and
(ii)
Vienna {\it versus} Mathematica and Maple (as already mentioned).

\begin{table}
\caption{Welch's $t$-test for the ``book-stack'' tests.}\label{tab:10}
\begin{center}
\begin{tabular}
[c]{ c c c c c }\hline\hline
$p$-value & Mathematica & Quantis & Vienna & $\pi$\\\hline
Maple  & 0.0535 & \bf{0.0436} & 0.974 & 0.3412\\
Mathematica &  & \bf{0.0009} & \bf{0.0283} & 0.1551\\
Quantis  &  &  & \bf{0.0368} & \bf{0.0054}\\
Vienna &  &  &  & 0.2690\\\hline\hline
\end{tabular}
\end{center}
\end{table}

The Kolmogorov-Smirnov test results are presented in Table~\ref{tab:11},
where no significant differences are detected.
The Shapiro-Wilk test results are presented in Table~\ref{tab:12}.
Since there is no clear
pattern of normality for the data, the application of Welch's $t$-test is not
appropriate.

\begin{table}
\caption{Kolmogorov-Smirnov test for the algorithmic information theory tests.}\label{tab:11}
 \begin{center}
 \begin{tabular}
[c]{ c c c c c }
\hline\hline
Kolmogorov-Smirnov test $p$-values & Mathematica & Quantis & Vienna & $\pi$ \\\hline
Maple  & 0.7591 & 0.4005 & 0.7591 & 0.7591\\
Mathematica &  & 0.7591 & 0.7591 & 0.7591\\
Quantis &  &  & 0.4005 & 0.7591\\
Vienna &  &  &  & 0.9883\\\hline\hline
\end{tabular}
\end{center}
\end{table}

\begin{table}
\caption{Shapiro-Wilk test for the algorithmic information theory tests.}\label{tab:12}
\begin{center}
\begin{tabular}
[c]{ c c c c c c }\hline\hline
Shapiro-Wilk test & Maple & Mathematica & Quantis  & Vienna  & $\pi$\\\hline
$p$-value & 0.0696 & \bf{0.0363} &
0.4378 & 0.6963 & 0.4315\\\hline\hline
\end{tabular}
\end{center}
\end{table}

\subsubsection{Test based on random walks}

The Kolmogorov-Smirnov test results are presented in Table~\ref{tab:13}.

\begin{table}
\caption{Kolmogorov-Smirnov test for the random walk tests.}\label{tab:13}
 \begin{center}
 \begin{tabular}
[c]{ c c c c c }
\hline\hline
Kolmogorov-Smirnov test $p$-values & Mathematica & Quantis & Vienna & $\pi$ \\\hline
Mathematica & 0.1678 & \bf{0.0123} & 0.4175 & 0.0525\\
Quantis &  & $\mathbf{< 10^{-4}}$ & \bf{0.0021} & 0.1678\\
Vienna &  &  & 0.0525 & $\mathbf{< 10^{-4}}$\\
$\pi$ &  &  &  & \bf{0.0002}\\\hline\hline
\end{tabular}
\end{center}
\end{table}

Statistically significant differences are identified for:
(i) Quantis {\it versus} all other sources (Maple, Mathematica, Vienna and $\pi$);
(ii) Vienna {\it versus} Mathematica, Vienna (as already
mentioned) and  $\pi$;  and
(iii) Maple {\it versus} $\pi$.

Note that quantum strings move farther away from the starting point than the pseudorandom strings; i.e.,
$\langle \text{Vienna} \rangle >\langle \text{Quantis} \rangle >\langle \text{Maple} \rangle >\langle \text{Mathematica} \rangle >\langle  \pi \rangle $.

It was quite natural to double-check the conclusion  ``Quantis and Vienna don't
exhibit significant difference.'' Hence we run the Shapiro-Wilk test which concludes
that normality is not rejected; cf. Table~\ref{tab:14}.

\begin{table}
\caption{Shapiro-Wilk test for the random walk tests.}\label{tab:14}
\begin{center}
\begin{tabular}
[c]{ c c c c c c }\hline\hline
Shapiro-Wilk test & Maple & Mathematica & Quantis & Vienna  & $\pi$\\\hline
$p$-value & 0.2006 & 0.9268 & 0.5464 & 0.8888 &
0.9577\\\hline\hline
\end{tabular}
\end{center}
\end{table}

Next, we apply the Welch's $t$-test for the comparison of means. The
results are given in Table~\ref{tab:15}.
Significant differences between the means are identified for the
following sources:
(i) Quantis {\it versus} all other sources (Maple, Quantis, Vienna, $\pi$);
(ii)
Vienna {\it versus} Mathematica), Quantis (as already mentioned) and $\pi$;
(iii) Maple {\it versus} $\pi$.

\begin{table}
\caption{Welch's $t$-tests for the random walk tests.}\label{tab:15}
\begin{center}
\begin{tabular}
[c]{ c c c c c }\hline\hline
$p$-value & Mathematica & Quantis & Vienna & $\pi$\\\hline
Maple & 0.06961 & \bf{0.0013} & 0.1409 & \bf{0.0119}\\
Mathematica &  & $\mathbf{< 10^{-4}}$ & \bf{0.0007} & \bf{0.0435}\\
Quantis &  &  & \bf{0.0143} & $\mathbf{< 10^{-4}}$\\
Vienna &  &  &  & \bf{0.0001}\\\hline\hline
\end{tabular}
\end{center}
\end{table}

\if01


\subsection[blabla]{T-information-based tests~\footnote{This section was written by U.~Speidel.}}

The tests described  in this section were performed on the files {\tt random{\it n}.bits} with {\tt\it n} between 0 and 7. These are referred to as ``string 0'' to ``string 7'' below.

For hardware limitations, the strings were divided into non-overlapping substrings of fixed length. Three lengths were investigated: 500,000, 1 million and 5 million bytes, yielding 8 sets of 1073, 536, and 107 substrings each. These lengths were chosen as a compromise between size of string (more reliable estimate through T-information) and size of sample population required for meaningful statistical tests.

Titchener's {\em T-information} $I_T$ (see~\cite{titchener-96}) has been computed for each substring, yielding $3 \times 8$ sets of $I_T$ values. For each set, the mean and standard deviation have been computed (rounded to nearest integer):

500,000 bytes:\\
{\small
\begin{tabular}{ l| r r r r r r r r }
\hline
String & 0 & 1 & 2 & 3 & 4 & 5 & 6 & 7 \\
\hline
Average T-info (nats) & 1821867
 & 1821836
 & 1821854
 & 1821849
 & 1821919
 & 1821814
 & 1821895
 & 1821867 \\
\hline
Stdev (nats) & 1527 & 1509 & 1490 & 1493 & 1494 & 1503 & 1545 \\
\hline
\end{tabular}}

1 million bytes:\\
{\small
\begin{tabular}{ l| r r r r r r r r }
\hline
String & 0 & 1 & 2 & 3 & 4 & 5 & 6 & 7 \\
\hline
Average T-info (nats) & 3513081 & 3513190 & 3513137 & 3513139 & 3513223 & 3513162 & 3513130 & 3513190 \\
\hline
Stdev (nats) & 2293 & 2387 & 2443 & 2383 & 2408 & 2357 & 2364 & 2162 \\
\hline
\end{tabular}}

5 million bytes:\\
{\small
\begin{tabular}{ l| r r r r r r r r }
\hline
String & 0 & 1 & 2 & 3 & 4 & 5 & 6 & 7 \\
\hline
Average T-info (nats) & 17681157 & 17682078 & 17681777 & 17681433 & 17680964 & 17681728 & 17680266 & 17681522\\
\hline
Stdev (nats) & 5698 & 5525 & 5681 & 5314 & 5312 & 5537 & 5449 & 6676\\
\hline
\end{tabular}}

Note that, with respect to the standard deviation, string 7 is found at the extreme ends of the range in both cases, but not at the same end.

For each pair of original strings, the associated substring distributions of $I_T$ were compared statistically using a $t$-test and an f-test.

For the 500,000 byte substrings, neither the $t$-test nor the f-test yield any significant differences between the original strings using an 0.1 confidence threshold.

The $t$-test for the 28 possible pairings of 1 million byte substring distributions also shows no significant difference between the distributions.

However, distributions of string 7 clearly differ from those of all other strings except string 0 in the f-test for the same pairings if a
confidence threshold of 0.05 is applied:

1 million bytes:\\
{\small
\begin{tabular}{ l| r r r r r r r r }
\hline
f-test String & 0 & 1 & 2 & 3 & 4 & 5 & 6 \\
\hline
vs.\ String 7& 0.17  & 0.022 & 0.0048 & 0.024 & 0.013 & 0.046 & 0.039 \\
\hline
\end{tabular}}

None of the other pairings produce f-test values anywhere near the 0.05 threshold: the lowest confidence value is 0.14 between strings~2 and 0,
the next lowest 0.26 -- results one would have to expect in 21 pairings of samples from the same distribution.

For the 5 million byte substrings, the $t$-test shows two pairings below the 0.05 threshold and one very close to it, all associated with string~6,
in comparison with strings 1 (0.017), 2 (0.048), and 5 (0.053) respectively. The next lowest value, 0.11, is also associated with string~6.
This indicates that string 6 may be another outlier.

The f-test once again flags string~7 as different, with 3 values under the 0.05 threshold and a further three below the 0.1 threshold (and the remaining
value just at 0.1):

5 million bytes:\\
{\small
\begin{tabular}{ l| r r r r r r r r }
\hline
f-test String & 0 & 1 & 2 & 3 & 4 & 5 & 6 \\
\hline
vs.\ String 7& 0.10  & 0.053 & 0.098 & 0.020 & 0.019 & 0.056 & 0.038 \\
\hline
\end{tabular}}
The pairings affected are the same as for the 1~million byte substrings. No other pairing not involving string~7 exhibits f-test values of less than 0.47.

For quality control, the tests above were then repeated, using only the first and last half, respectively, of the substrings of each string. That is, each distribution was divided into two, using the first/last 268 substrings of 1~million bytes and the first/last 53/54 substrings of 5 million bytes. These were again subjected to $t$-tests and f-tests. String~7 is the one clear outlier that can be identified with some confidence. Its f-test is incompatible with all other strings except 0 at the 0.05 threshold. It is also flagged as different in a significant number of $t$-tests for length 5 million bytes.
\fi

\if01
\subsection{Tests based on T-information}
U. Speidel~\cite{Speidel} has used T-information~\cite{titchener-96}
to perform a limited set
of tests on eight random sequences (the \emph{anonymous strings}
were, in this order, Vienna1, Mathematica1, Pi, Quantis1, Clib,
Quantis2, Mathematica2, and Vienna2).  The strings of length $2^{32}$
bits were divided into non-overlapping substrings of fixed
length. Three lengths were investigated: 500,000, 1 million and 5
million bytes, yielding eight sets of 1073, 536, and 107 substrings
each. These lengths were chosen as a compromise between size of
string (more reliable estimate through T-information) and size of
sample population required for meaningful statistical
tests~\footnote{Also note that the
memory limitations of Spiedel's software prevented us from
calculating the T-information value for the whole string of length
$2^{32}$.}. The Vienna
strings (0 and 7) appear to be the clear outlier(s)---
in particular, string 7 can be identified to be from a different source
with respect to strings 3, 4 and 6 at the 0.05 confidence level, and with
respect to 1, 2, and 5 at the 0.1 confidence level  (see
Table~\ref{tab:ftests5M}).

\begin{table}[ht]
\caption{F-tests on full distributions of 5M byte snippets of eight random
sequences.}\label{tab:ftests5M}

\

\centerline{\fbox{\includegraphics[width=5.8in]{Ulrich5Morig8}}}
\end{table}
\fi

\section{Conclusions}

\label{conclusion}

Our aim was to experimentally study the possibility of distinguishing between
quantum sources of randomness---recently proved to be theoretically incomputable---and  some well-known computable
sources of  pseudo-randomness. The experimental approach is based on algorithmic information theory
which provides characterizations of algorithmic random sequences in terms of the degrees of
randomness of their finite prefixes.
In this theory the degree of incompressibility of a string is measured with the prefix-complexity, which, unfortunately, is incomputable.
Fortunately, there are computable, but weaker properties than
incompressibility which can be tested on prefixes.
Of course, such a property is necessary but not sufficient, so the (degree of) absence of the property is significant.

We have performed
tests of randomness on pseudo-random strings (finite sequences) of length $2^{32}$ generated with software
(Mathematica, Maple), which are  cyclic (so, strongly computable), the bits of $\pi$, which is computable, but not cyclic, and strings produced by quantum measurements  (with the commercial device Quantis and  by the Vienna IQOQI group).

It is important to emphasize that our aim was to find tests capable of  distinguishing computable
from incomputable sources of ``randomness'' by examining (long, but) finite prefixes of infinite sequences.
Such differences are guaranteed to exist by the result in Ref~\cite{2008-cal-svo}, but,
because computability is an asymptotic property,
there was no guarantee that finite tests can ``pick'' them in the prefixes we have analyzed~\footnote{Many inconclusive tests have been discarded.}.

With these {\em privisos}, our empirical randomness tests indicate quantitative differences between computable
and incomputable sources of ``randomness;''
more specifically:
\begin{enumerate}
\item
pseudo-random strings perform very well on Borel normality---in fact, too well (some overestimate by more than 2\%
of length), while the Vienna strings---which have not been post-processed---indicate
deviations from Borel normality for test strings of small length (up to length 4);

\item in computing Shannon's entropy for our sequences we observe that
the average seems to be the same for all sources. However,
the Vienna sources clearly show a much flatter ``Bell curve'' around
its median; the Quantis results are somewhat peculiar in that the median is
clearly not centered within the 50\% percentile of the entropies
(indicating a skewed Bell curve) and the Mathematica sequences have
a few outliers with large entropy;

\item
in the random walk test quantum random sources (both Vienna and Quantis)
seem to move farther away from the starting point than the pseudo-generators.
\item
the test based on the correctness of probabilistic tests of primality is  more ``utilitarian,''
as the metric reflects the length of the sample ``random'' string necessary for the Solovay-Strassen algorithm
to reach the correct answer; overall, quantum random generators appear to be different from pseudo-random
 generators; with the Vienna strings emerging as the clear outlier
(in all tests with various degrees of confidence);

\item
the behavior of $\pi$ (computable, but not cyclic) is interesting: in tests 1, 4
and 5 the results are closer to Mathematica and Maple, in tests 2 and 3 the
results for $\pi$ stands out (above) of all others in the direction of possibly
being ``more'' random (according to these test metrics).
\end{enumerate}

The statistical analysis of the randomness tests shows that the Borel normality test is the best test
(from our collection) for detecting and differentiating between the computable and incomputable random sources;
the random walk test and the ``book-stack''  follow in efficiency.
  The Shannon test and the test based on probabilistic primality behavior~\cite{calude:02} do not produce
 statistically significant results.
In the first case the reason may come from the fact  that averages are the same for all samples. In the second case  the reason  may be due to the fact that the test is based solely on the behavior of algorithmic random strings and not on a specific property of randomness.

 The pair of tests based on Borel normality and random walks seem to address complimentary properties helping to distinguish well between
 computable and incomputable sources of ``randomness.''
Pseudo-random strings perform better  than quantum strings for the Borel normality test.
 One could speculate that pseudo-randomness incorporates
the ``human'' perception of randomness, which is strongly associated with uniform distribution;
in contrast, quantum randomness has no such bias.
Quantum random bits  tend to take a longer time to reach
``uniform distribution''---which is an asymptotical property---than pseudo-random strings.

Our analysis indicate normality of the (finite) quantum sequences for longer test strings,
but violations of normality for a few small length test strings (up to length 4).
Notice that for finite sequences of quantum or other origin,
normality needs not be satisfied for all test strings;
hence the derivations cannot be taken as a clear signal of a violation of Borel normality
stemming, say, from a lack of independence.
With these caveats, a conceivable (speculative and  by no means necessary) physical explanation of this violation of normality
for test strings of small length would be that, due to photon
(Bose-Einstein) statistics and the Hanbury-Brown-Twiss effect
(``photon bunching;'' i.e., the tendency of photons to arrive in identical states), independence and thus Borel
normality might be violated for ``small" groups of data.
In this line of thought,  for larger sequences a sort of ``late randomness'' becomes visible,
as the short-term correlations disappear in time.
In contrast, for the random walk test, which addresses a global type of behaviour rather than
a local one, quantum strings perform better: they tend to  move farther away from the starting point.

A few more caveats are in order.
As expected, our results indicate some tendencies only.  As this is a first attempt to experimentally distinguish computable from incomputable sources
of ``randomness,'' much more work is necessary to understand those differences.
  New tests should be designed to reflect the asymptotic differences. We may work with  longer strings of bits to trespass the
cyclicality of the pseudo-random generators~\footnote{Borel normality obviously  fails for
longer strings.}. We suggest that there may be different types of ``quantum randomness''
corresponding to different forms of quantum indeterminism (e.g.,
entanglement, Bell's theorem, Kochen-Specker theorem).
Finally, our experimental results clearly cannot, and do not aim, to ``prove'' in any formal way  the
superiority of quantum random generators over the best pseudo-random ones for practical applications;
the only superiority is asymptotic, and resides in the differences between computable and incomputable sources proven in Ref~\cite{2008-cal-svo}.

\section*{Acknowledgements}
We are grateful to  Thomas Jennewein and Anton Zeilinger for providing us with the quantum random bits produced at the University of Vienna by the Vienna IQOQI group, for the description of their method, critical comments and  interest in this research.

We thank: a) Alastair Abbott, Hector Zenil and Boris Ryabko for interesting comments, b)
Ulrich Speidel for his tests for which some partial results have been reported in our extended paper~\cite{CDMTCS372},
c) Stefan Wegenkittl for critical comments of various drafts of this paper and his suggestions  to exclude  some tests.

Cristian Calude gratefully  acknowledges the support of  the Hood Foundation (Fellowship Grant 2008--2009) and the  Technical University of Vienna (where his work was done during his visits in 2008 and 2009).
Karl Svozil  gratefully  acknowledges support of the
Centre for Discrete Mathematics and Theoretical Computer Science (CDMTCS) at the University of Auckland, as well as of the Ausseninstitut of the Vienna University of Technology.


\begin{thebibliography}{100}
\newcommand{\enquote}[1]{``#1''}
\expandafter\ifx\csname url\endcsname\relax
  \def\url#1{{#1}}\fi
\expandafter\ifx\csname urlprefix\endcsname\relax\def\urlprefix{}\fi

\bibitem{frank}
P.~Frank, {\em Das Kausalgesetz und seine Grenzen\/} (Springer, Vienna, 1932),
  {E}nglish translation in Ref.~\cite{franke}.

\bibitem{poincare14}
H.~Poincar{\'{e}}, {\em Wissenschaft und Hypothese\/} (Teubner, Leipzig, 1914).

\bibitem{Diacu96}
F.~Diacu, \enquote{The Solution of the N-body Problem,} The Mathematical
  Intelligencer {\bf 18}, 66--70 (1996).
\newline http://dx.doi.org/10.1007/BF03024313

\bibitem{Sundman12}
K.~E. Sundman, \enquote{Memoire sur le probl{\`{e}}me de trois corps,} Acta
  Mathematica {\bf 36}, 105--179 (1912).

\bibitem{Wang91}
Q.~D. Wang, \enquote{The global solution of the $N$-body problem,} Celestial
  Mechanics {\bf 50}, 73--88 (1991).
\newline http://dx.doi.org/10.1007/BF00048987

\bibitem{Wang01}
Q.~D. Wang, \enquote{Power Series Solutions and Integral Manifold of the n-body
  Problem,} Regular \& Chaotic Dynamics {\bf 6}, 433--442 (2001).
\newline http://dx.doi.org/10.1070/RD2001v006n04ABEH000187

\bibitem{eckmann1}
J.-P. Eckmann and D.~Ruelle, \enquote{Ergodic theory of chaos and strange
  attractors,} Reviews of Modern Physics {\bf 57}, 617--656 (1985).
\newline http://dx.doi.org/10.1103/RevModPhys.57.617

\bibitem{Diacu96-ce}
F.~Diacu and P.~Holmes, {\em Celestial Encounters - the Origins of Chaos and
  Stability\/} (Princeton University Press, Princeton, 1996).

\bibitem{born-55}
M.~Born, \enquote{{I}st die klassische {M}echanik tats{\"{a}}chlich
  deterministisch?} Physikalische Bl{\"{a}}tter {\bf 11}, 49--54 (1955),
  {E}nglish translation ``Is classical mechanics in fact deterministic?''
  Reprinted in Ref.~\cite[p.~78-83]{born-69}.

\bibitem{Note1}
Recall Einstein's {\protect \it dictum} in a letter to Born, dated
  December~12th, 1926~\cite [p.~113]{born-69}, ``In any case I am convinced
  that he [[the Old One]] does not throw dice.'' (In German: ``Jedenfalls bin
  ich {\"{u}}berzeugt, dass der [[Alte]] nicht w{\"{u}}rfelt.'').

\bibitem{pauli-probaphysics}
W.~Pauli, \enquote{{W}ahrscheinlichkeit und {P}hysik,} Dialectica {\bf 8},
  112--124 (1954), {E}nglish translation in
  Ref.~\cite[pp.~43-48]{pauli-philosophy}.
\newline http://dx.doi.org/10.1111/j.1746-8361.1954.tb01125.x

\bibitem{zeil-05_nature_ofQuantum}
A.~Zeilinger, \enquote{The message of the quantum,} Nature {\bf 438}, 743
  (2005).
\newline http://dx.doi.org/10.1038/438743a

\bibitem{ch-schw-78}
G.~J. Chaitin and J.~T. Schwartz, \enquote{A note on monte carlo primality
  tests and algorithmic information theory,} Communications on Pure and Applied
  Mathematics {\bf 31}, 521--527 (1978).
\newline http://dx.doi.org/10.1002/cpa.3160310407

\bibitem{Granville-92}
A.~Granville, \enquote{Primality testing and Carmichael numbers,} Notices of
  the American Mathematical Society {\bf 39}, 696--700 (1992).
\newline http://www.dms.umontreal.ca/~andrew/PDF/Notices1.pdf

\bibitem{PhysRevE.69.055702}
S.~Mertens and H.~Bauke, \enquote{Entropy of pseudo-random-number generators,}
  Physical Review E {\bf 69}, 055\,702 (2004).
\newline http://dx.doi.org/10.1103/PhysRevE.69.055702

\bibitem{v-neumann-50}
N.~C. Metropolis, G.~Reitweisner, and J.~von Neumann, \enquote{Statistical
  Treatment of Values of First 2000 decimal digits of $e$ and of $\pi$
  Calculated on the {ENIAC},} Mathematical Tables and Other Aids to Computation
  {\bf 4}, 109--111 (1950), reprinted in {\sl John von Neumann, Collected
  Works, (Vol. V)}, A. H. Traub, editor, MacMillan, New York, 1963, p. 765.

\bibitem{Marsaglia-68}
G.~Marsaglia, \enquote{random numbers fall mainly in the planes,} Proceedings
  of the National Academy of Sciences of the United States of America (PNAS)
  {\bf 61}, 25--28 (1968).
\newline http://www.pnas.org/content/61/1/25.full.pdf

\bibitem{DBLP:journals/ibmrd/Pickover91}
C.~A. Pickover, \enquote{Picturing randomness on a graphics supercomputer,} IBM
  Journal of Research and Development {\bf 35}, 227--230 (1991).
\newline http://www.research.ibm.com/journal/rd/351/ibmrd3501a2S.pdf

\bibitem{Bowman1995315}
R.~L. Bowman, \enquote{Evaluating pseudo-random number generators,} Computers
  \& Graphics {\bf 19}, 315--324 (1995).
\newline http://dx.doi.org/10.1016/0097-8493(94)00158-U

\bibitem{von-neumann1}
J.~von Neumann, \enquote{Various Techniques Used in Connection With Random
  Digits,} National Bureau of Standards Applied Math Series {\bf 12}, 36--38
  (1951), reprinted in {\sl John von Neumann, Collected Works, (Vol. V)}, A. H.
  Traub, editor, MacMillan, New York, 1963, p. 768--770.

\bibitem{diaconis:211}
P.~Diaconis, S.~Holmes, and R.~Montgomery, \enquote{Dynamical Bias in the Coin
  Toss,} SIAM Review {\bf 49}, 211--235 (2007).
\newline http://dx.doi.org/10.1137/S0036144504446436

\bibitem{csw:prg}
C.~S. Wallace, \enquote{Physically random generator,} Computer Systems Science
  \& Engineering {\bf 5}, 82--88 (1990).

\bibitem{0022-3735-3-8-303}
C.~H. Vincent, \enquote{The generation of truly random binary numbers,} Journal
  of Physics E: Scientific Instruments {\bf 3}, 594--598 (1970), corrigendum in
  Ref.~\cite{0022-3735-3-10-528}.
\newline http://dx.doi.org/10.1088/0022-3735/3/8/303

\bibitem{Agnew-87}
G.~B. Agnew, \enquote{Random sources for cryptographic systems,} in {\em
  Advances in Cryptology - EUROCRYPT'87\/}, A.~Adamatzky, ed.  (Springer,
  Berlin, 1987), pp. 77--82.
\newline http://dsns.csie.nctu.edu.tw/research/crypto/HTML/PDF/E87/77.PDF

\bibitem{Aware}
{Aware Electronics Corp.}
\newline http://www.aw-el.com

\bibitem{Araneus}
{Araneus Information Systems Oy, Araneus Alea I True Random Number Generator}.
\newline http://www.araneus.fi/products-alea-eng.html

\bibitem{comscire}
{ComScire - Quantum World Corp}.
\newline http://www.comscire.com

\bibitem{LavaRnd}
{LavaRnd Random Number Generator}.
\newline http://lavarnd.org

\bibitem{stipcevic045104}
M.~Stip\v{c}evi\'{c} and B.~M. Rogina, \enquote{Quantum random number generator
  based on photonic emission in semiconductors,} Review of Scientific
  Instruments {\bf 78}, 045\,104 (2007).
\newline http://dx.doi.org/10.1063/1.2720728

\bibitem{dynes:031109}
J.~F. Dynes, Z.~L. Yuan, A.~W. Sharpe, and A.~J. Shields, \enquote{A high
  speed, postprocessing free, quantum random number generator,} Applied Physics
  Letters {\bf 93}, 031\,109 (2008).
\newline http://dx.doi.org/10.1063/1.2961000

\bibitem{Wayne-09}
M.~A.Wayne, E.~R. Jeffrey, G.~M. Akselrod, and P.~G. Kwiat, \enquote{Photon
  arrival time quantum random number generation,} Journal of Modern Optics {\bf
  56}, 516--516 (2009).
\newline http://dx.doi.org/10.1080/09500340802553244

\bibitem{Ma:05}
H.-Q. Ma, Y.~Xie, and L.-A. Wu, \enquote{Random number generation based on the
  time of arrival of single photons,} Applied Optics {\bf 44}, 7760--7763
  (2005).
\newline http://dx.doi.org/10.1364/AO.44.007760

\bibitem{rand-55}
{The RAND Corporation}, {\em A Million Random Digits with 100,000 Normal
  Deviates Free Press Publishers\/} (Knolls Atomic Power Lab. Report KAPL-3147,
  Glencoe, Illinois, 1955), the data digits are obtainable {\em via}
  http://www.rand.org/pubs/monograph\_reports/2005/digits.txt.zip, the
  introduction {\it via}
  http://www.rand.org/pubs/monograph\_reports/MR1418/index2.html.
\newline http://www.rand.org/pubs/monograph\_reports/MR1418/

\bibitem{Note2}
According to The RAND Corporation's disclosure, ``The random digits in this
  book were produced by re-randomization of a basic table generated by an
  electronic roulette wheel. Briefly, a random frequency pulse source,
  providing on the average about 100,000 pulses per second, was gated about
  once per second by a constant frequency pulse. Pulse standardization circuits
  passed the pulses through a 5-place binary counter. In principle the machine
  was a 32-place roulette wheel which made, on the average, about 3000
  revolutions per trial and produced one number per second. A binary-to-decimal
  converter was used which converted 20 of the 32 numbers (the other twelve
  were discarded) and retained only the final digit of two-digit numbers; this
  final digit was fed into an IBM punch to produce finally a punched card table
  of random digits.''.

\bibitem{PhysRevLett.54.1023}
R.~J. Cook and H.~J. Kimble, \enquote{Possibility of Direct Observation of
  Quantum Jumps,} Physical Review Letters {\bf 54}, 1023--1026 (1985).
\newline http://dx.doi.org/10.1103/PhysRevLett.54.1023

\bibitem{er-put:85}
T.~Erber and S.~Putterman, \enquote{Randomness of quantum mechanics: nature's
  ultimate cryptogram?} Nature {\bf 318}, 41--43 (1985).
\newline http://dx.doi.org/10.1038/318041a0

\bibitem{erber-95}
T.~Erber, \enquote{Testing the Randomness of Quantum Mechanics: Nature's
  Ultimate Cryptogram?} in {\em Annals of the New York Academy of Sciences.
  {V}olume 755 Fundamental Problems in Quantum Theory\/}, D.~M. Greenberger and
  A.~Zeilinger, eds.  (Springer, Berlin, Heidelberg, New York, 1995), Vol. 755,
  pp. 748--756.
\newline http://dx.doi.org/10.1111/j.1749-6632.1995.tb39016.x

\bibitem{knight-86}
P.~L. Knight, R.~Loudon, and D.~T. Pegg, \enquote{Quantum jumps and atomic
  cryptograms,} Nature {\bf 323}, 608--609 (1986).
\newline http://dx.doi.org/10.1038/323608a0

\bibitem{schmidt:462}
H.~Schmidt, \enquote{Quantum-Mechanical Random-Number Generator,} Journal of
  Applied Physics {\bf 41}, 462--468 (1970).
\newline http://dx.doi.org/10.1063/1.1658698

\bibitem{walker-hotbits}
J.~Walker, \enquote{HotBits Hardware,}  (1986-2009).
\newline http://www.fourmilab.ch/hotbits/hardware3.html

\bibitem{stipcevic4442}
M.~Stip\v{c}evi\'{c}, \enquote{Fast nondeterministic random bit generator based
  on weakly correlated physical events,} Review of Scientific Instruments {\bf
  75}, 4442--4449 (2004).
\newline http://dx.doi.org/10.1063/1.1809295

\bibitem{svozil-qct}
K.~Svozil, \enquote{The quantum coin toss---Testing microphysical
  undecidability,} Physics Letters A {\bf 143}, 433--437 (1990).
\newline http://dx.doi.org/10.1016/0375-9601(90)90408-G

\bibitem{rarity-94}
J.~G. Rarity, M.~P.~C. Owens, and P.~R. Tapster, \enquote{Quantum Random-number
  Generation and Key Sharing,} Journal of Modern Optics {\bf 41}, 2435--2444
  (1994).
\newline http://dx.doi.org/10.1080/09500349414552281

\bibitem{zeilinger:qct}
T.~Jennewein, U.~Achleitner, G.~Weihs, H.~Weinfurter, and A.~Zeilinger,
  \enquote{A Fast and Compact Quantum Random Number Generator,} Review of
  Scientific Instruments {\bf 71}, 1675--1680 (2000).
\newline http://dx.doi.org/10.1063/1.1150518

\bibitem{stefanov-2000}
A.~Stefanov, N.~Gisin, O.~Guinnard, L.~Guinnard, and H.~Zbinden,
  \enquote{Optical quantum random number generator,} Journal of Modern Optics
  {\bf 47}, 595--598 (2000).
\newline http://dx.doi.org/10.1080/095003400147908

\bibitem{0256-307X-21-10-027}
M.~Hai-Qiang, W.~Su-Mei, Z.~Da, C.~Jun-Tao, J.~Ling-Ling, H.~Yan-Xue, and
  W.~Ling-An, \enquote{A Random Number Generator Based on Quantum Entangled
  Photon Pairs,} Chinese Physics Letters {\bf 21}, 1961--1964 (2004).
\newline http://dx.doi.org/10.1088/0256-307X/21/10/027

\bibitem{wang:056107}
P.~X. Wang, G.~L. Long, and Y.~S. Li, \enquote{Scheme for a quantum random
  number generator,} Journal of Applied Physics {\bf 100}, 056\,107 (2006).
\newline http://dx.doi.org/10.1063/1.2338830

\bibitem{fiorentino:032334}
M.~Fiorentino, C.~Santori, S.~M. Spillane, R.~G. Beausoleil, and W.~J. Munro,
  \enquote{Secure self-calibrating quantum random-bit generator,} Physical
  Review A (Atomic, Molecular, and Optical Physics) {\bf 75}, 032\,334 (2007).
\newline http://dx.doi.org/10.1103/PhysRevA.75.032334

\bibitem{svozil-2009-howto}
K.~Svozil, \enquote{Three criteria for quantum random-number generators based
  on beam splitters,} Physical Review A (Atomic, Molecular, and Optical
  Physics) {\bf 79}, 054\,306 (2009).
\newline http://dx.doi.org/10.1103/PhysRevA.79.054306

\bibitem{Kwon:09}
O.~Kwon, Y.-W. Cho, and Y.-H. Kim, \enquote{Quantum random number generator
  using photon-number path entanglement,} Applied Optics {\bf 48}, 1774--1778
  (2009).
\newline http://dx.doi.org/10.1364/AO.48.001774

\bibitem{jammer:89}
M.~Jammer, {\em The Conceptual Development of Quantum Mechanics. 2nd edition.
  The History of Modern Physics, 1800-1950; v. 12\/} (American Institute of
  Physics, New York, 1989).

\bibitem{jammer1}
M.~Jammer, {\em The Philosophy of Quantum Mechanics\/} (John Wiley \& Sons, New
  York, 1974).

\bibitem{feynman-law}
R.~P. Feynman, {\em The Character of Physical Law\/} (MIT Press, Cambridge, MA,
  1965).

\bibitem{fuchs-peres}
C.~A. Fuchs and A.~Peres, \enquote{Quantum theory needs no `Interpretation´,}
  Physics Today {\bf 53}, 70--71 (2000), further discussions of and reactions
  to the article can be found in the September issue of Physics Today, {\it
  53}, 11-14 (2000).
\newline http://www.aip.org/web2/aiphome/pt/vol-53/iss-9/p11.html and
  http://www.aip.org/web2/aiphome/pt/vol-53/iss-9/p14.html

\bibitem{clauser-talkvie}
J.~Clauser, \enquote{Early History of {B}ell´s Theorem,} in {\em Quantum
  (Un)speakables. {F}rom {B}ell to Quantum Information\/} (Springer, Berlin,
  2002), pp. 61--96.

\bibitem{born-26-1}
M.~Born, \enquote{Zur {Q}uantenmechanik der {S}to{\ss}vorg{\"{a}}nge,}
  Zeitschrift f{\"{u}}r Physik {\bf 37}, 863--867 (1926).
\newline http://dx.doi.org/10.1007/BF01397477

\bibitem{born-26-2}
M.~Born, \enquote{{Q}uantenmechanik der {S}to{\ss}vorg{\"{a}}nge,} Zeitschrift
  f{\"{u}}r Physik {\bf 38}, 803--827 (1926).
\newline http://dx.doi.org/10.1007/BF01397184

\bibitem{pauli:58}
W.~Pauli, \enquote{{D}ie allgemeinen {P}rinzipien der {W}ellenmechanik,} in
  {\em {H}andbuch der {P}hysik. {B}and {V}, {T}eil 1. {P}rinzipien der
  {Q}uantentheorie {I}\/}, S.~Fl{\"{u}}gge, ed.  (Springer, Berlin,
  G{\"{o}}ttingen and Heidelberg, 1958), pp. 1--168.

\bibitem{peres222}
A.~Peres, \enquote{Unperformed experiments have no results,} American Journal
  of Physics {\bf 46}, 745--747 (1978).
\newline http://dx.doi.org/10.1119/1.11393

\bibitem{mermin-93}
N.~D. Mermin, \enquote{Hidden variables and the two theorems of {J}ohn {B}ell,}
  Reviews of Modern Physics {\bf 65}, 803--815 (1993).
\newline http://dx.doi.org/10.1103/RevModPhys.65.803

\bibitem{wheeler-Zurek:83}
J.~A. Wheeler and W.~H. Zurek, {\em Quantum Theory and Measurement\/}
  (Princeton University Press, Princeton, 1983).

\bibitem{Note3}
{ ``Vom Standpunkt unserer Quantenmechanik gibt es keine Gr\"o\ss e, die im
  {\protect \em Einzelfalle} den Effekts eines Sto\ss es kausal festlegt; aber
  auch in der Erfahrung haben wir keinen Anhaltspunkt daf\"ur, da\ss ~ es
  innere Eigenschaften der Atome gibt, die einen bestimmten Sto\ss erfolg
  bedingen. Sollen wir hoffen, sp\"ater solche Eigenschaften [[$\protect \dots
  $]] zu entdecken und im Einzelfalle zu bestimmen? Oder sollen wir glauben,
  dass die \"Ubereinstimmung von Theorie und Erfahrung in der Unf\"ahigkeit,
  Bedingungen f\"ur den kausalen Ablauf anzugeben, eine pr\"astabilisierte
  Harmonie ist, die auf der Nichtexistenz solcher Bedingungen beruht? Ich
  selber neige dazu,die Determiniertheit in der atomaren Welt aufzugeben.'' }.

\bibitem{Note4}
{ ``Die Bewegung der Partikel folgt Wahrscheinlichkeitsgesetzen, die
  Wahrscheinlichkeit selbst aber breitet sich im Einklang mit dem Kausalgesetz
  aus. [Das hei\ss t, da\ss ~ die Kenntnis des Zustandes in allen Punkten in
  einem Augenblick die Verteilung des Zustandes zu allen sp{\"a}teren Zeiten
  festlegt.]'' }.

\bibitem{mermin-07}
N.~D. Mermin, {\em Quantum Computer Science\/} (Cambridge University Press,
  Cambridge, 2007).
\newline http://people.ccmr.cornell.edu/~mermin/qcomp/CS483.html

\bibitem{svozil-2008-ql}
K.~Svozil, \enquote{Contexts in quantum, classical and partition logic,} in
  {\em Handbook of Quantum Logic and Quantum Structures\/}, K.~Engesser, D.~M.
  Gabbay, and D.~Lehmann, eds.  (Elsevier, Amsterdam, 2009), pp. 551--586.
\newline http://arxiv.org/abs/quant-ph/0609209

\bibitem{v-neumann-49}
J.~von Neumann, {\em Mathematische Grundlagen der Quantenmechanik\/} (Springer,
  Berlin, 1932), {E}nglish translation in Ref.~\cite{v-neumann-55}.

\bibitem{kochen1}
S.~Kochen and E.~P. Specker, \enquote{The Problem of Hidden Variables in
  Quantum Mechanics,} Journal of Mathematics and Mechanics (now Indiana
  University Mathematics Journal) {\bf 17}, 59--87 (1967), reprinted in
  Ref.~\cite[pp. 235--263]{specker-ges}.
\newline http://dx.doi.org/10.1512/iumj.1968.17.17004

\bibitem{neumark-54}
M.~A. Neumark, \enquote{Principles of quantum theory,} in {\em {S}owjetische
  {A}rbeiten zur {F}unktionalanalysis. {B}eiheft zur {S}owjetwissenschaft\/},
  K.~Matthes, ed.  (Gesellschaft f{\"{u}}r Deutsch-Sowjetische Freundschaft,
  Berlin, 1954), Vol.~44, pp. 195--273.

\bibitem{halmos-vs}
P.~R. Halmos, {\em Finite-dimensional vector spaces\/} (Springer, New York,
  Heidelberg, Berlin, 1974).

\bibitem{hkwz}
T.~J. Herzog, P.~G. Kwiat, H.~Weinfurter, and A.~Zeilinger,
  \enquote{Complementarity and the quantum eraser,} Physical Review Letters
  {\bf 75}, 3034--3037 (1995).
\newline http://dx.doi.org/10.1103/PhysRevLett.75.3034

\bibitem{greenberger2}
D.~M. Greenberger and A.~YaSin, \enquote{``{H}aunted'' measurements in quantum
  theory,} Foundation of Physics {\bf 19}, 679--704 (1989).
\newline http://dx.doi.org/10.1007/BF00731905

\bibitem{wiesner}
S.~Wiesner, \enquote{Conjugate coding,} SIGACT News {\bf 15}, 78--88 (1983).
\newline http://dx.doi.org/10.1145/1008908.1008920

\bibitem{Note5}
See also the later patents at Refs.~\cite {dultz-98,dultz-99}, as well as at
  Refs.~\cite {Ribordy-04,Ribordy-06}.

\bibitem{zeilinger-epr-98}
G.~Weihs, T.~Jennewein, C.~Simon, H.~Weinfurter, and A.~Zeilinger,
  \enquote{Violation of {B}ell's Inequality under Strict {E}instein Locality
  Conditions,} Phys. Rev. Lett. {\bf 81}, 5039--5043 (1998).
\newline http://dx.doi.org/10.1103/PhysRevLett.81.5039

\bibitem{murnaghan}
F.~D. Murnaghan, {\em The Unitary and Rotation Groups\/} (Spartan Books,
  Washington, D.C., 1962).

\bibitem{Mandel-Ou1987118}
Z.~Ou, C.~Hong, and L.~Mandel, \enquote{Relation between input and output
  states for a beam splitter,} Optics Communications {\bf 63}, 118--122 (1987).
\newline http://dx.doi.org/10.1016/0030-4018(87)90271-9

\bibitem{green-horn-zei}
D.~M. Greenberger, M.~A. Horne, and A.~Zeilinger, \enquote{Multiparticle
  interferometry and the superposition principle,} Physics Today {\bf 46},
  22--29 (1993).

\bibitem{zeilinger:882}
A.~Zeilinger, \enquote{General properties of lossless beam splitters in
  interferometry,} American Journal of Physics {\bf 49}, 882--883 (1981).
\newline http://dx.doi.org/10.1119/1.12387

\bibitem{svozil-2004-analog}
K.~Svozil, \enquote{Noncontextuality in multipartite entanglement,} J. Phys. A:
  Math. Gen. {\bf 38}, 5781--5798 (2005).
\newline http://dx.doi.org/10.1088/0305-4470/38/25/013

\bibitem{Note6}
``To cite a human example, for simplicity, in tossing a coin it is probably
  easier to make two consecutive tosses independent than to toss heads with
  probability exactly one-half. If independence of successive tosses is
  assumed, we can reconstruct a 50--50 chance out of even a badly biased coin
  by tossing twice. If we get heads-heads or tails-tails, we reject the tosses
  and try again. If we get heads-tails (or tails-heads), we accept the result
  as heads (or tails).''.

\bibitem{elias-72}
P.~Elias, \enquote{The Efficient Construction of an Unbiased Random Sequence,}
  Ann. Math. Statist. {\bf 43}, 865--870 (1972).
\newline http://dx.doi.org/10.1214/aoms/1177692552

\bibitem{PeresY-1992}
Y.~Peres, \enquote{Iterating {V}on {N}eumann's procedure for extracting random
  bits,} The Annals of Statistics {\bf 20}, 590--597 (1992).
\newline http://www.jstor.org/stable/2242181

\bibitem{dichtl-2007}
M.~Dichtl, \enquote{Bad and Good Ways of Post-processing Biased Physical Random
  Numbers,} in {\em Fast Software Encryption. Lecture Notes in Computer Science
  Volume 4593/2007\/}, A.~Biryukov, ed.  (Springer, Berlin and Heidelberg,
  2007), pp. 137--152, 14th International Workshop, FSE 2007, Luxembourg,
  Luxembourg, March 26-28, 2007, Revised Selected Papers.
\newline http://dx.doi.org/10.1007/978-3-540-74619-5\_9

\bibitem{Lacharme-2008}
P.~Lacharme, \enquote{Post-Processing Functions for a Biased Physical Random
  Number Generator,} in {\em Fast Software Encryption. Lecture Notes in
  Computer Science Volume 5086/2008\/}, K.~Nyberg, ed.  (Springer, Berlin and
  Heidelberg, 2008), pp. 334--342, 15th International Workshop, FSE 2008,
  Lausanne, Switzerland, February 10-13, 2008, Revised Selected Papers.
\newline http://dx.doi.org/10.1007/978-3-540-71039-4\_21

\bibitem{chau}
J.~C. Garrison and R.~Y. Chiao, {\em Quantum Optics\/} (Oxford, Oxford, 2008).

\bibitem{knight-qo}
C.~Gerry and P.~L. Knigh, {\em Introductory Quantum Optics\/} (Cambridge
  University Press, Cambridge, UK, 2005).

\bibitem{schrodinger}
E.~Schr{\"{o}}dinger, \enquote{Die gegenw{\"{a}}rtige {S}ituation in der
  {Q}uantenmechanik,} Naturwissenschaften {\bf 23}, 807--812, 823--828,
  844--849 (1935), {E}nglish translation in Ref.~\cite{trimmer} and in
  Ref.~\cite[pp. 152-167]{wheeler-Zurek:83}.
\newline http://dx.doi.org/10.1007/BF01491891,
  \\http://dx.doi.org/10.1007/BF01491914,
  \\http://dx.doi.org/10.1007/BF01491987

\bibitem{Note7}
In its {\protect \em White Paper on Random Numbers Generation using Quantum
  Physics}~\cite {Quantis}, {\protect \em id Quantique} on p.~7 (in the caption
  to Fig.~1) announces that its {\protect \em Quantis} device uses a light
  emitting diode, while at the same time (top of p.~7) pointing out that the
  monitoring of a Zener diode is problematic: ``Formally the evolution of these
  generators is not random, but just very complex. One could say that
  determinism is hidden behind complexity.''.

\bibitem{e-f-moore}
E.~F. Moore, \enquote{Gedanken-Experiments on Sequential Machines,} in {\em
  Automata Studies\/}, C.~E. Shannon and J.~McCarthy, eds.  (Princeton
  University Press, Princeton, 1956), pp. 129--153.

\bibitem{wright}
R.~Wright, \enquote{Generalized urn models,} Foundations of Physics {\bf 20},
  881--903 (1990).
\newline http://dx.doi.org/10.1007/BF01889696

\bibitem{svozil-2001-eua}
K.~Svozil, \enquote{Logical equivalence between generalized urn models and
  finite automata,} International Journal of Theoretical Physics {\bf 44},
  745--754 (2005).
\newline http://dx.doi.org/10.1007/s10773-005-7052-0

\bibitem{svozil-2005-ln1e}
K.~Svozil, \enquote{Staging quantum cryptography with chocolate balls,}
  American Journal of Physics {\bf 74}, 800--803 (2006).
\newline http://dx.doi.org/10.1119/1.2205879

\bibitem{misra:756}
B.~Misra and E.~C.~G. Sudarshan, \enquote{The Zeno's paradox in quantum
  theory,} Journal of Mathematical Physics {\bf 18}, 756--763 (1977).
\newline http://dx.doi.org/10.1063/1.523304

\bibitem{Heisenberg-27}
W.~Heisenberg, \enquote{{{\"{U}}ber den anschaulichen Inhalt der
  quantentheoretischen Kinematik und Mechanik},} Zeitschrift für Physik {\bf
  43}, 172--198 (1927), english translation in Ref.~\cite[pp.
  62-84]{wheeler-Zurek:83}.
\newline http://dx.doi.org/10.1007/BF01397280

\bibitem{vonNeumann:1927:WAQ}
J.~von Neumann, \enquote{{Wahrscheinlichkeitstheoretischer Aufbau der
  Quantenmechanik}. ({German}) [{Probabilistic} structure of quantum
  mechanics],} {Nachrichten von der Gesellschaft der Wissenschaften zu
  G{\"o}ttingen} {\bf 1}, 245--272 (1927), reprinted in
  \cite[Paper~10]{Taub:1961:JNCa}.
\newline http://www.digizeitschriften.de/main/dms/img/?IDDOC=465854

\bibitem{dirac}
P.~A.~M. Dirac, {\em The Principles of Quantum Mechanics\/} (Oxford University
  Press, Oxford, 1930).

\bibitem{Note8}
{ ``Bei der Unbestimmtheit einer Eigenschaft eines Systems bei einer bestimmten
  Anordnung (bei einem bestimmten Zustand eines Systems) vernichtet jeder
  Versuch, die betreffende Eigenschaft zu messen, (mindestens teilweise) den
  Einflu\ss ~ der fr{\"u}heren Kenntnisse vom System auf die (eventuell
  statistischen) Aussagen {\"u}ber sp{\"a}tere m{\"o}gliche Messungsergebnisse.
  [[$\protect \dots $]] So m{\"u}ssen, um den Ort eines Teilchens zu bestimmen
  und um seinen Impuls zu bestimmen, {\protect \em einander ausschlie\ss ende
  Versuchsanordnungen benutzt werden.} [[$\protect \dots $]] Die Beeinflussung
  des Systems durch den Messaparat f{\"u}r den Impuls (Ort) ist eine solche,
  da\ss ~ innerhalb der durch die Ungenauigkeitsrelationen gegebenen Grenzen
  die Benutzbarkeit der fr{\"u}heren Orts- (Impuls-) Kenntnis f{\"u}r die
  Voraussagbarkeit der Ergebnisse sp{\"a}terer Orts- (Impuls-) Messungen
  verlorengegangen ist. Wenn aus diesem Grunde die Benutzbarkeit {\protect \em
  eines} klassischen Begriffes in einem ausschlie\ss enden Verh{\"a}ltnis zu
  einem {\protect \em anderen} steht, nennen wir diese beiden Begriffe (z.B.
  Orts- und Impulskoordinaten eines Teilchens) mit Bohr {\protect \em
  komplement{\"a}r.} [[$\protect \dots $]] Man wird sehen, dass diese
  ``Komplementarit{\"a}t'' kein Analogon in der klassischen Gastheorie besitzt,
  die ja auch mit statistischen Gesetzm\"a\ss igkeiten operiert. Diese Theorie
  enth{\"a}lt n{\"a}mlich nicht die erst durch die Endlichkeit des
  Wirkungsquantums geltend werdende Aussage, da\ss ~ durch Messungen an einem
  System die durch fr{\"u}here Messungen gewonnenen Kenntnisse {\"u}ber das
  System unter Umst{\"a}nden notwendig verlorengehen m{\"u}ssen, d.h. nicht
  mehr verwertet werden k{\"o}nnen.'' }.

\bibitem{epr}
A.~Einstein, B.~Podolsky, and N.~Rosen, \enquote{Can quantum-mechanical
  description of physical reality be considered complete?} Physical Review {\bf
  47}, 777--780 (1935).
\newline http://dx.doi.org/10.1103/PhysRev.47.777

\bibitem{svozil-2006-omni}
K.~Svozil, \enquote{Quantum Scholasticism: On Quantum Contexts,
  Counterfactuals, and the Absurdities of Quantum Omniscience,} Information
  Sciences {\bf 179}, 535--541 (2009).
\newline http://dx.doi.org/10.1016/j.ins.2008.06.012

\bibitem{svozil-2006-uniquenessprinciple}
K.~Svozil, \enquote{Are simultaneous Bell measurements possible?} New Journal
  of Physics {\bf 8}, 39 (2006).
\newline http://dx.doi.org/10.1088/1367-2630/8/3/039

\bibitem{Gleason}
A.~M. Gleason, \enquote{Measures on the closed subspaces of a {H}ilbert space,}
  Journal of Mathematics and Mechanics (now Indiana University Mathematics
  Journal) {\bf 6}, 885--893 (1957).
\newline http://dx.doi.org/10.1512/iumj.1957.6.56050

\bibitem{pitowsky:218}
I.~Pitowsky, \enquote{Infinite and finite Gleason's theorems and the logic of
  indeterminacy,} Journal of Mathematical Physics {\bf 39}, 218--228 (1998).
\newline http://dx.doi.org/10.1063/1.532334

\bibitem{rich-bridge}
F.~Richman and D.~Bridges, \enquote{A constructive proof of {G}leason's
  theorem,} Journal of Functional Analysis {\bf 162}, 287--312 (1999).
\newline http://dx.doi.org/10.1006/jfan.1998.3372

\bibitem{r:dvur-93}
A.~Dvure{\v{c}}enskij, {\em {G}leason's Theorem and Its Applications\/} (Kluwer
  Academic Publishers, Dordrecht, 1993).

\bibitem{bell}
J.~S. Bell, \enquote{On the {E}instein {P}odolsky {R}osen paradox,} Physics
  {\bf 1}, 195--200 (1964), reprinted in Ref.~\cite[pp.
  403-408]{wheeler-Zurek:83} and in \cite[pp. 14-21]{bell-87}.

\bibitem{hey-red}
P.~Heywood and M.~L.~G. Redhead, \enquote{Nonlocality and the
  {K}ochen-{S}pecker Paradox,} Foundations of Physics {\bf 13}, 481--499
  (1983).
\newline http://dx.doi.org/10.1007/BF00729511

\bibitem{ghz}
D.~M. Greenberger, M.~A. Horne, and A.~Zeilinger, \enquote{Going beyond Bell's
  theorem,} in {\em Bell's Theorem, Quantum Theory, and Conceptions of the
  {U}niverse\/}, M.~Kafatos, ed.  (Kluwer Academic Publishers, Dordrecht,
  1989), pp. 73--76.

\bibitem{specker-60}
E.~Specker, \enquote{{D}ie {L}ogik nicht gleichzeitig entscheidbarer
  {A}ussagen,} Dialectica {\bf 14}, 239--246 (1960), reprinted in
  Ref.~\cite[pp. 175--182]{specker-ges}; {E}nglish translation: {\it The logic
  of propositions which are not simultaneously decidable}, Reprinted in
  Ref.~\cite[pp. 135-140]{hooker}.
\newline http://dx.doi.org/10.1111/j.1746-8361.1960.tb00422.x

\bibitem{ZirlSchl-65}
N.~Zierler and M.~Schlessinger, \enquote{Boolean embeddings of orthomodular
  sets and quantum logic,} Duke Mathematical Journal {\bf 32}, 251--262 (1965).

\bibitem{Alda}
V.~Alda, \enquote{On\/ {\rm 0-1} measures for projectors I,} Aplik. mate. {\bf
  25}, 373--374 (1980).

\bibitem{Alda2}
V.~Alda, \enquote{On\/ {\rm 0-1} measures for projectors II,} Aplik. mate. {\bf
  26}, 57--58 (1981).

\bibitem{kamber64}
F.~Kamber, \enquote{Die {S}truktur des {A}ussagenkalk{\"{u}}ls in einer
  physikalischen {T}heorie,} Nachr. Akad. Wiss. G{\"{o}}ttingen {\bf 10},
  103--124 (1964).

\bibitem{kamber65}
F.~Kamber, \enquote{Zweiwertige {W}ahrscheinlichkeitsfunktionen auf
  orthokomplement{\"{a}}ren {V}erb{\"{a}}nden,} Mathematische Annalen {\bf
  158}, 158--196 (1965).

\bibitem{peres-91}
A.~Peres, \enquote{Two simple proofs of the {K}ochen-{S}pecker theorem,}
  Journal of Physics A: Mathematical and General {\bf 24}, L175--L178 (1991),
  reprinted in Ref.~\cite[pp. 186-200]{peres}.
\newline http://dx.doi.org/10.1088/0305-4470/24/4/003

\bibitem{svozil-tkadlec}
K.~Svozil and J.~Tkadlec, \enquote{Greechie diagrams, nonexistence of measures
  in quantum logics and {K}ochen--{S}pecker type constructions,} Journal of
  Mathematical Physics {\bf 37}, 5380--5401 (1996).
\newline http://dx.doi.org/10.1063/1.531710

\bibitem{cabello-96}
A.~Cabello, J.~M. Estebaranz, and G.~Garc{\'{i}}a-Alcaine,
  \enquote{{B}ell-{K}ochen-{S}pecker theorem: A proof with 18 vectors,} Physics
  Letters A {\bf 212}, 183--187 (1996).
\newline http://dx.doi.org/10.1016/0375-9601(96)00134-X

\bibitem{cabello:210401}
A.~Cabello, \enquote{Experimentally Testable State-Independent Quantum
  Contextuality,} Physical Review Letters {\bf 101}, 210\,401 (2008).
\newline http://dx.doi.org/10.1103/PhysRevLett.101.210401

\bibitem{bohr-1949}
N.~Bohr, \enquote{Discussion with {E}instein on epistemological problems in
  atomic physics,} in {\em {A}lbert {E}instein: Philosopher-Scientist\/}, P.~A.
  Schilpp, ed.  (The Library of Living Philosophers, Evanston, Ill., 1949), pp.
  200--241.
\newline http://www.emr.hibu.no/lars/eng/schilpp/Default.html

\bibitem{bell-66}
J.~S. Bell, \enquote{On the Problem of hidden variables in quantum mechanics,}
  Reviews of Modern Physics {\bf 38}, 447--452 (1966), reprinted in
  Ref.~\cite[pp. 1-13]{bell-87}.
\newline http://dx.doi.org/10.1103/RevModPhys.38.447

\bibitem{redhead}
M.~Redhead, {\em Incompleteness, Nonlocality, and Realism: A Prolegomenon to
  the Philosophy of Quantum Mechanics\/} (Clarendon Press, Oxford, 1990).

\bibitem{Note9}
Other schemes to circumvent the quantum value indefiniteness are through
  probabilities defined via paradoxical set decompositions~\cite
  {pitowsky-82,pitowsky-83} or by considering certain dense subsets of scarcely
  interlinked quantum contexts~\cite {meyer:99}.

\bibitem{svozil:040102}
K.~Svozil, \enquote{Proposed direct test of a certain type of noncontextuality
  in quantum mechanics,} Physical Review A (Atomic, Molecular, and Optical
  Physics) {\bf 80}, 040\,102 (2009).
\newline http://dx.doi.org/10.1103/PhysRevA.80.040102

\bibitem{wjswz-98}
G.~Weihs, T.~Jennewein, C.~Simon, H.~Weinfurter, and A.~Zeilinger,
  \enquote{Violation of {B}ell's Inequality under Strict {E}instein Locality
  Conditions,} Phys. Rev. Lett. {\bf 81}, 5039--5043 (1998).
\newline http://dx.doi.org/10.1103/PhysRevLett.81.5039

\bibitem{shimony2}
A.~Shimony, \enquote{Controllable and uncontrollable non-locality,} in {\em
  Proceedings of the International Symposium on the Foundations of Quantum
  Mechanics\/}, S.~K. {\it et al.}, ed.,  pp. 225--230 (1984), see also J.
  Jarrett, {\sl Bell's Theorem, Quantum Mechanics and Local Realism}, Ph. D.
  thesis, Univ. of Chicago, 1983; {\sl Nous}, {\bf 18}, 569 (1984).

\bibitem{2008-cal-svo}
C.~S. Calude and K.~Svozil, \enquote{Quantum Randomness and Value
  Indefiniteness,} Advanced Science Letters {\bf 1}, 165--168 (2008).
\newline
  http://www.ingentaconnect.com/content/asp/asl/2008/00000001/00000002/art00004

\bibitem{peres}
A.~Peres, {\em Quantum Theory: Concepts and Methods\/} (Kluwer Academic
  Publishers, Dordrecht, 1993).

\bibitem{svozil-ql}
K.~Svozil, {\em Quantum Logic\/} (Springer, Singapore, 1998).

\bibitem{calude:02}
C.~Calude, {\em Information and Randomness---An Algorithmic Perspective\/}
  (Springer, Berlin, 2002), 2nd edn.

\bibitem{Rukhin-nist}
A.~Rukhin, J.~Soto, J.~Nechvatal, M.~Smid, E.~Barker, S.~Leigh, M.~Levenson,
  M.~Vangel, D.~Banks, A.~Hekert, J.~Dray, and S.~Vo, {\em A Statistical Test
  Suite for Random and Pseudorandom Number Generators for Cryptographic
  Applications. NIST Special Publication 800-22\/} (National Institute of
  Standards and Technology (NIST), 2001).
\newline http://csrc.nist.gov/groups/ST/toolkit/rng/documents/SP800-22b.pdf

\bibitem{MRG}
\enquote{Mathematica random generator,} .
\newline
  http://reference.wolfram.com/mathematica/tutorial/RandomNumberGeneration.html

\bibitem{MAPLE}
\enquote{Maple random generator,} .
\newline
  http://www.maplesoft.com/applications/app\_center\_view.aspx?AID=2027\&CID=4%
\&SCID=9

\bibitem{Quantis}
id~Quantique, \enquote{The Quantis Quantum Random Number Generator,}
  (2001-2009).
\newline http://www.idquantique.com/products/files/quantis-whitepaper.pdf

\bibitem{Vienna}
{IQOQI Group Vienna}, personal communication.

\bibitem{pi}
Y.~Kanada and D.~Takahashi, \enquote{Calculation of $\pi$ up to 4,294,960,000
  decimal digits,}  (1995).
\newline ftp://pi.super-computing.org

\bibitem{Note10}
For the curious, our ten pairs of deleted digits were $\protect \{0,1\protect
  \},\protect \{0,5\protect \},\protect \{1,6\protect \},\protect \{2,3\protect
  \},\protect \{2,7\protect \},\protect \{3,8\protect \},\protect \{4,5\protect
  \},\protect \{4,9\protect \},\protect \{6,7\protect \},$ and $\protect
  \{8,9\protect \}$.

\bibitem{CDMTCS372}
C.~S. Calude, M.~J. Dinneen, M.~Dumitrescu, and K.~Svozil, \enquote{How Random
  Is Quantum Randomness? (Extended Version),} Report {CDMTCS}-372, Centre for
  Discrete Mathematics and Theoretical Computer Science, University of
  Auckland, Auckland, New Zealand (2009).
\newline http://www.cs.auckland.ac.nz/CDMTCS/researchreports/372crismjdkarl.pdf

\bibitem{borel:09}
E.~Borel, \enquote{Les probabilit\'{e}s d\'{e}nombrables et leurs applications
  arithm\'{e}tiques,} Rendiconti del Circolo Matematico di Palermo (1884 -
  1940) {\bf 27}, 247--271 (1909).
\newline http://dx.doi.org/10.1007/BF03019651

\bibitem{ZL:1978}
 .

\bibitem{DBLP:conf/dlt/Calude93}
C.~Calude, \enquote{{B}orel Normality and Algorithmic Randomness,} in {\em
  Developments in Language Theory\/}, G.~Rozenberg and A.~Salomaa, eds.  (World
  Scientific, Singapore, 1994), pp. 113--129.

\bibitem{Wyner}
A.~D. Wyner, \enquote{Shannon Lecture: Typical Sequences and All That: Entropy,
  Pattern Matching, and Data Compression,} IEEE Information Theory Society
  (1994).
\newline
  http://www.itsoc.org/people/awards-and-honors/claude-e.-shannon-award/Wyner9%
4Shannon.pdf/view

\bibitem{MR2099021}
B.~Y. Ryabko and A.~I. Pestunov, \enquote{``{B}ook stack'' as a new statistical
  test for random numbers,} Problemy Peredachi Informatsii {\bf 40}, 73--78
  (2004).

\bibitem{MR2162569}
B.~Y. Ryabko and V.~A. Monarev, \enquote{Using information theory approach to
  randomness testing,} J. Statist. Plann. Inference {\bf 133}, 95--110 (2005).

\bibitem{solovay:84}
R.~Solovay and V.~Strassen, \enquote{A Fast {M}onte-{C}arlo Test for
  Primality,} SIAM Journal on Computing {\bf 6}, 84--85 (1977), corrigendum in
  Ref.~\cite{solovay:118}.
\newline http://dx.doi.org/10.1137/0206006

\bibitem{Note11}
In fact, every ``decent'' Monte Carlo simulation algorithm in which tests are
  chosen according to an algorithmic random string produces a result which is
  not only true with high probability, but {\protect \it rigorously
  correct}~\cite {MR757602}.

\bibitem{Note12}
There are 1,401,644 Carmichael numbers in the interval $[1, 10^{18}]$.

\bibitem{Pinch}
R.~G. Pinch, \enquote{The Carmichael numbers up to $10^{16}$,}  (1998).
\newline http://arxiv.org/abs/math.NT/9803082

\bibitem{Pinch07}
R.~G. Pinch, \enquote{The Carmichael numbers up to $10^{21}$,} in {\em
  Proceedings of Conference on Algorithmic Number Theory 2007. {TUCS} General
  Publication No 46\/}  pp. 129--131 (2007).
\newline http://tucs.fi/publications/attachment.php?fname=G46.pdf

\bibitem{Conover}
W.~J. Conover, {\em Practical Nonparametric Statistics\/} (John Wiley \& Sons,
  New York, 1999).

\bibitem{Shapiro-Wilk}
S.~S. Shapiro and M.~B. Wilk, \enquote{An analysis of variance test for
  normality (complete samples),} Biometrika {\bf 52}, 591--611 (2005).
\newline http://dx.doi.org/10.1093/biomet/52.3-4.591

\bibitem{Welch}
B.~L. Welch, \enquote{The generalization of ``Student's'' problem when several
  different population variances are involved,} Biometrika {\bf 34} (1947).
\newline http://dx.doi.org/10.1093/biomet/34.1-2.28

\bibitem{rproject}
T.~R. Foundation, \enquote{The {R} Project for Statistical Computing, Version
  2.10.0,} Http://www.r-project.org.
\newline http://www.r-project.org

\bibitem{Note13}
Many inconclusive tests have been discarded.

\bibitem{Note14}
Borel normality obviously fails for longer strings.

\bibitem{franke}
P.~Frank and {R. S. Cohen (Editor)}, {\em The Law of Causality and its Limits
  (Vienna Circle Collection)\/} (Springer, Vienna, 1997).

\bibitem{born-69}
M.~Born, {\em Physics in my generation\/} (Springer Verlag, New York, 1969),
  2nd edn.

\bibitem{pauli-philosophy}
W.~Pauli, {\em Writings on physics and philosophy\/} (Springer Verlag, Berlin,
  New York, 1994), ed. by Charles Paul Enz and Karl {von Meyenn}.

\bibitem{0022-3735-3-10-528}
C.~H. Vincent, \enquote{The generation of truly random binary numbers,} Journal
  of Physics E: Scientific Instruments {\bf 3}, 832 (1970).
\newline http://dx.doi.org/10.1088/0022-3735/3/10/528

\bibitem{v-neumann-55}
J.~von Neumann, {\em Mathematical Foundations of Quantum Mechanics\/}
  (Princeton University Press, Princeton, 1955).

\bibitem{specker-ges}
E.~Specker, {\em Selecta\/} (Birkh{\"{a}}user Verlag, Basel, 1990).

\bibitem{dultz-98}
W.~Dultz and E.~Hildebrandt, \enquote{Optical random-check generator based on
  the individual photon statistics at the optical beam divider. ({G}erman:
  {O}ptischer {Z}ufallsgenerator basierend auf der {E}inzelphotonenstatistik am
  optischen {S}trahlteiler),}  (1999), patent Pub. No.: WO/1998/016008,
  International Application No.: PCT/EP1997/005082, Publication Date:
  16.04.1998, International Filing Date: 17.09.1997, Chapter 2 Demand Filed:
  23.04.1998, IPC: H03K 3/84 (2006.01).
\newline http://www.wipo.int/pctdb/en/wo.jsp?wo=1998016008

\bibitem{dultz-99}
W.~Dultz, G.~Dultz, E.~Hildebrandt, and H.~Schmitzer, \enquote{Method for
  generating a random number on a quantum-mechanics basis and random generator.
  ({G}erman: {V}erfahren zur {E}rzeugung einer {Z}ufallszahl auf
  quantenmechanischer {G}rundlage und {Z}ufallsgenerator,}  (1999), patent Pub.
  No.: WO/1999/066641, International Application No.: PCT/EP1999/003689,
  Publication Date: 23.12.1999, International Filing Date: 28.05.1999, IPC:
  G06F 7/58 (2006.01), H03K 3/84 (2006.01).
\newline http://www.wipo.int/pctdb/en/wo.jsp?wo=1999066641

\bibitem{Ribordy-04}
G.~Ribordy and O.~Guinnard, \enquote{Method and apparatus for generating true
  random numbers by way of a quantum optics process,}  (2004), patent
  Application number: 10/919,573, Publication number: US 2005/0071400 A1,
  Filing date: Aug 17, 2004, U.S. Classification 708250000, International
  Classification G06F001/02.
\newline http://www.google.com/patents?id=eQqXAAAAEBAJ

\bibitem{Ribordy-06}
G.~Ribordy and O.~Guinnard, \enquote{Method and apparatus for generating true
  random numbers by way of a quantum optics process,}  (2006), patent
  Application number: 11/422,704, Publication number: US 2007/0127718 A1,
  Filing date: Jun 7, 2006, U.S. Classification 380256000.
\newline http://www.google.com/patents?id=BUmiAAAAEBAJ

\bibitem{trimmer}
J.~D. Trimmer, \enquote{The present situation in quantum mechanics: a
  translation of {S}chr{\"{o}}dinger's ``cat paradox'',} Proceedings of the
  American Philosophical Society {\bf 124}, 323--338 (1980), reprinted in
  Ref.~\cite[pp. 152-167]{wheeler-Zurek:83}.
\newline http://www.tu-harburg.de/rzt/rzt/it/QM/cat.html

\bibitem{Taub:1961:JNCa}
J.~von Neumann, {\em {John von Neumann}: Collected Works: {Volume I}: {Logic},
  Theory of Sets and Quantum Mechanics\/} (Pergamon, New York, NY, 1961).

\bibitem{bell-87}
J.~S. Bell, {\em Speakable and Unspeakable in Quantum Mechanics\/} (Cambridge
  University Press, Cambridge, 1987).

\bibitem{hooker}
C.~A. Hooker, {\em The Logico-Algebraic Approach to Quantum Mechanics. {V}olume
  {I}: Historical Evolution\/} (Reidel, Dordrecht, 1975).

\bibitem{pitowsky-82}
I.~Pitowsky, \enquote{Resolution of the {E}instein-{P}odolsky-{R}osen and
  {B}ell paradoxes,} Physical Review Letters {\bf 48}, 1299--1302 (1982).
\newline http://dx.doi.org/10.1103/PhysRevLett.48.1299

\bibitem{pitowsky-83}
I.~Pitowsky, \enquote{Deterministic model of spin and statistics,} Physical
  Review D {\bf 27}, 2316--2326 (1983).
\newline http://dx.doi.org/10.1103/PhysRevD.27.2316

\bibitem{meyer:99}
D.~A. Meyer, \enquote{Finite precision measurement nullifies the
  {K}ochen-{S}pecker theorem,} Physical Review Letters {\bf 83}, 3751--3754
  (1999).
\newline http://dx.doi.org/10.1103/PhysRevLett.83.3751

\bibitem{solovay:118}
R.~Solovay and V.~Strassen, \enquote{Erratum: A Fast Monte-Carlo Test for
  Primality,} SIAM Journal on Computing {\bf 7}, 118 (1978).
\newline http://dx.doi.org/10.1137/0207009

\bibitem{MR757602}
C.~Calude and M.~Zimand, \enquote{A relation between correctness and randomness
  in the computation of probabilistic algorithms,} Internat. J. Comput. Math.
  {\bf 16}, 47--53 (1984).

\end{thebibliography}

\end{document}